\documentclass[traditabstract,twocolumn]{aa} 

\usepackage{epsfig,graphicx,natbib}
\usepackage{longtable}

\begin{document}

\title{Molecules at $z$=0.89:}
\subtitle{A 4-mm-rest-frame absorption-line survey toward PKS~1830$-$211}

\author{S. Muller \inst{1}
\and A. Beelen \inst{2}
\and M. Gu\'elin \inst{3,4}
\and S. Aalto \inst{1}
\and J. H. Black \inst{1}
\and F. Combes \inst{5}
\and S. J. Curran \inst{6}
\and P. Theule \inst{7}
\and S. N. Longmore \inst{8}
}

\institute{ Onsala Space Observatory, SE 439-92, Onsala, Sweden
\and Institut d'Astrophysique Spatiale, B\^at. 121, Universit\'e Paris-Sud, 91405 Orsay Cedex, France
\and Institut de Radioastronomie Millim\'etrique, 300, rue de la piscine, 38406 St Martin d'H\`eres, France 
\and Ecole Normale Sup\'erieure/LERMA, 24 rue Lhomond, 75005 Paris, France
\and Observatoire de Paris, LERMA, CNRS, 61 Av. de l'Observatoire, 75014 Paris, France
\and School of Physics, University of New South Wales, Sydney NSW 2052, Australia
\and Physique des interactions ioniques et mol\'eculaires, Universit\'e de Provence, Centre de Saint J\'er\^ome, 13397 Marseille Cedex 20, France
\and ESO, Karl-Schwarzschild-Str. 2, 85748 Garching, Germany
}

\date {Received  / Accepted}

\titlerunning{Molecules at $z$=0.89}
\authorrunning{Muller et al. 2011}

\abstract{ 
We present the results of a 7~mm spectral survey of molecular absorption lines originating in the disk of a $z$=0.89
spiral galaxy located in front of the quasar PKS~1830$-$211. Our survey was performed with the Australia Telescope
Compact Array and covers the frequency interval 30--50 GHz, corresponding to the rest-frame frequency interval
57--94 GHz. A total of 28 different species, plus 8 isotopic variants, were detected toward the south-west
absorption region, located about 2~kpc from the center of the $z$=0.89 galaxy, 
which therefore has the largest number of detected molecular species of any extragalactic object so far.
The results of our rotation diagram analysis show that the rotation temperatures are close to the cosmic microwave
background temperature of 5.14~K that we expect to measure at $z$=0.89, whereas the kinetic temperature is one
order of magnitude higher, indicating that the gas is subthermally excited.
The molecular fractional abundances are found to be in-between those in typical Galactic diffuse and
translucent clouds, and clearly deviate from those observed in the dark cloud TMC~1 or in the Galactic center
giant molecular cloud Sgr~B2. The isotopic ratios of carbon, nitrogen, oxygen, and silicon deviate significantly
from the solar values, which can be linked to the young age of the $z$=0.89 galaxy and a release of nucleosynthesis
products dominated by massive stars.
Toward the north-east absorption region, where the extinction and column density of gas is roughly one order of magnitude
lower than toward the SW absorption region, only a handful of molecules are detected. Their relative abundances are comparable
to those in Galactic diffuse clouds.
We also report the discovery of several new absorption components, with velocities spanning between $-$300 and
+170~km\,s$^{-1}$.
Finally, the line centroids of several species (e.g., CH$_3$OH, NH$_3$) are found to be significantly offset
from the average velocity. If caused by a variation in the proton-to-electron
mass ratio $\mu$ with redshift, these offsets yield an upper limit $|\frac{\Delta \mu}{\mu}|$$<$4$\times$$10^{-6}$, 
which takes into account the kinematical noise produced by the velocity dispersion measured from a large number
of molecular species.


}

\maketitle

\section{Introduction}

More than 150 molecules have been discovered in space, as a result of targeted investigations or spectral
surveys toward Galactic sources such as circumstellar envelopes of evolved stars (e.g. IRC+10216,
\citealt{cer00,pat11}), massive star-forming clouds (e.g. Sgr~B2, \citealt{num00}), or cold molecular
clouds (e.g. TMC-1, \citealt{ohi92,kai04}). The study of the rotational and vibrational spectra of these
molecules is particularly useful to probe the physical conditions and chemical content of the interstellar
gas. Most importantly, the large number of molecules found in various environments demonstrates the richness
and variety of interstellar chemistry.

As the sensitivity of millimeter instruments has increased, a large number of molecules have become
observable in galaxies in the Local Universe (see e.g. \citealt{wan04, mar06} for NGC4945 and NGC253,
respectively) and molecules are still detected at high redshifts (currently up to $z$=6.42), though
only a few species such as CO, HCN, HCO$^+$, and HNC (e.g. \citealt{wal03, wag05, rie06, gue07}).
Those detections are mostly related to ultra-luminous galaxies, which are potentially unrepresentative
of all galaxies at these epochs. Moreover, line emission from these high-$z$ galaxies is dominated by
the warmest and densest regions of the molecular component.

Observations of molecular absorption lines toward a bright radio continuum source allow rare molecular
species to be detected even in distant galaxies, where they cannot be observed in emission because of
distance dilution. This technique can thus be used to investigate the chemistry and its complexity at
high redshifts and trace the chemical enrichment history of the Universe. In addition, high-$z$ molecular
absorbers can serve as cosmological probes of the cosmic microwave background (CMB) temperature and have
attracted much interest in constraining variations in the fundamental constants of physics (see e.g.,
\citealt{hen09} and references therein).

Unfortunately, only a handful of high-redshift radio molecular absorbers \footnote{That is, not including
the H$_2$/CO absorbers detected from optical/UV spectroscopy, such as those discussed in \cite{not11}.}
have been discovered to date (see the review by \citealt{com08}), despite numerous searches (e.g.
\citealt{wik96a,dri96,mur03,cur04,cur06}, see also \citealt{cur11}). The molecular absorption originating
in the $z$=0.89 and $z$=0.68 intervening galaxies located in front of the quasars PKS~1830$-$211 ($z$=2.5)
and B~0218+357 ($z$=0.94) respectively, were discovered by \cite{wik95,wik96b} and have many similarities:
both absorbers appear to be (faint) nearly face-on spiral galaxies (\citealt{win02,yor05}); both galaxies
act as gravitational lenses and split the image of their background quasar into two main compact components
and an Einstein ring, as seen at radio wavelengths (\citealt{jau91,pat93}); they both harbor large column
densities of absorbing molecular gas ($N($H$_2)> 10^{22}$ cm$^{-2}$); and in both cases the radio continuum
of the background quasar is bright enough to permit sensitive observations with current millimeter
instruments. A dozen molecular species have indeed been successfully detected in these absorbers, as
well as absorption by their rare isotopologues (cf e.g., \citealt{wik95,wik96b,wik98,men99,mul06}).

\begin{figure}[h] \begin{center}
\includegraphics[width=5cm]{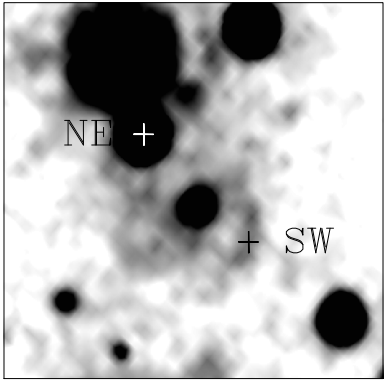}
\caption{Image of the $z$=0.89 galaxy obtained with the Hubble Space Telescope in the I band (see
also \citealt{mul06}). The positions of the SW and NE images of the quasar are indicated. The distance
between them is one arcsecond. Note the high extinction toward the SW image of the quasar, produced by
the interstellar medium of the intervening galaxy.}
\label{fig-hst}
\end{center} \end{figure}

The multiplicity of lensed continuum images of the background quasar makes it possible to explore as many
different lines of sight through the intervening galaxy. In the case of B~0218+357, molecular absorption
is detected only toward one image of the quasar (\citealt{men96,mul07}). Toward PKS~1830$-$211, molecular
absorption is detected, however, for both of the bright and compact images, located one arcsecond apart
on the north-east and south-west side, respectively, of the Einstein ring (\citealt{wik98,mul06}). The two
lines of sight intercept the disk of the $z$=0.89 galaxy (hereafter referred to as FG0.89, where FG
stands for foreground galaxy) on either side of its bulge, at galactocentric distances of $\sim$2~kpc
(to the SW) and $\sim$4~kpc (to the NE). An image of the galaxy, obtained with the Hubble Space Telescope
in the I band is shown in Fig.\ref{fig-hst}.

The absorption toward the SW image (hereafter FG0.89SW) has a large column of absorbing gas
($N$(H$_2)> 10^{22}$ cm$^{-2}$) and is taken as the reference 0 km\,s$^{-1}$ in velocity ($z$=0.88582,
\citealt{wik96b}). Remarkably, the full width at zero power of the HCO$^+$(2-1) line reaches nearly
100 km\,s$^{-1}$ for this component, although the galaxy is seen almost face-on. We note the relatively high
kinetic temperature of $\sim$80~K (\citealt{hen08}) and moderate density of a few times $10^3$ cm$^{-3}$
(\citealt{hen09}) toward FG0.89SW, which imply that the excitation of the rotational lines is subthermal
and mostly, but not necessarily totally, coupled with the CMB. Rotation temperatures derived for several
molecules are consistent with the value $T_{CMB}$=5.14~K at $z$=0.89 (\citealt{com97, men99, hen09}),
as predicted for an adiabatic expansion of the Universe.

The absorption toward the NE component (hereafter FG0.89NE) is located $\sim$147~km\,s$^{-1}$ blueward
and is narrower (FWHM$\sim$15~km\,s$^{-1}$), so that both velocity components can easily be resolved
kinematically. Nevertheless, the column of molecular gas toward the NE image is about two orders of
magnitude less than in FG0.89SW, and the corresponding absorption lines are significantly weaker.
Little is known about the excitation of and the physical conditions toward FG0.89NE. However, based
on the analysis of time variations in the NE absorption, \cite{mul08} argued that it should be caused
by a small number ($\le$5) of clouds of size $\sim$1 pc and of density a few 100 cm$^{-3}$, hence
resembling Galactic diffuse clouds.

Molecular absorption has also been detected in the galaxies hosting the radio sources B3~1504+377 at $z$=0.67
(in CO, HCO$^+$, HCN, and HNC; \citealt{wik96b}), PKS~1413+135 at $z$=0.25 (in CO, HCO$^+$, HCN, HNC,
and CN; \citealt{wik97}), and PKS~0132$-$097 at $z$=0.76 (although only in OH; \citealt{kan05}).
However, the weakness of the background radio continuum sources, narrower linewidths
and smaller column densities of absorbing gas than the two absorbers toward PKS~1830$-$211
and B~0218+357, ensures that it is difficult to perform a deep inventory of their molecular constituents with current
instruments.

In this paper, we present the first unbiased spectral-line survey toward a high-$z$ molecular absorber,
located at $z$=0.89 in front of the quasar PKS~1830$-$211. Observations, conducted with the Australian
Compact Array and covering the whole 7 mm band from 30 GHz to 50 GHz, are presented in \S.\ref{Obs}.
Our results are presented in \S.\ref{Results}, and then discussed in \S.\ref{Discussion}, in terms of
isotopic ratios (\S.\ref{section-isotopicratios}), the chemical composition relative to other objects
(\S.\ref{section-comparochemistry}), and constraints on the variation in fundamental constants of physics
(\S.\ref{section-velo}). Notes for the different molecular species detected in the survey are given in
appendix. 

\section{Observations and data reduction} \label{Obs}

Observations were carried out with the Australian Telescope Compact Array (ATCA) over three nights, on 2009
September 1 and 2 and on 2010 March 17 (see Table~\ref{tab-journal}). The array was in the EW352 and
H168 configuration, respectively, and we used the five inner antennas, discarding the sixth one located
4~km away as we obtained unsatisfactory pointing solutions and data quality. The maximum baseline length was
then up to 352~m and 192~m in either configuration, respectively. With this configuration, PKS~1830$-$211 was
mostly unresolved, and is further considered as a point-like source. The weather was clear in both runs.

For each tuning, the bright quasar 1923$-$293 was observed for several minutes to help us perform bandpass
calibration, and the pointing was checked and updated against PKS~1830$-$211, before integrating. We did not
observe gain calibrators, as PKS~1830$-$211 visibilities were self-calibrated and the continuum level
normalized to unity.

The Compact Array Broadband Backend (CABB) system was configured to provide a 1~MHz spectral resolution
per 2~GHz window with dual polarization. Two windows were used simultaneously, with central frequencies
set as reported in Table~\ref{tab-journal}. The whole 7~mm band (30--50~GHz) could then be covered in a
fairly small number of individual tunings. Each part of the band was covered in at least two tunings, in
order to prevent potential defectuous spectral channels. Some parts of the spectral band were indeed
affected by a series of bad channels, as can be seen for example in the final spectrum between 40 and
40.2~GHz in Fig.\ref{fig-fullsurvey}. Some frequency intervals were observed in more tunings, to increase
the sensitivity, for example at the frequencies of the deuterated species DCO$^+$ and DCN. The spectral
resolution of 1~MHz corresponds to $\sim$10~km\,s$^{-1}$ at 30~GHz and $\sim$6~km\,s$^{-1}$ at 50~GHz.

The data reduction was performed in two steps. We used the MIRIAD package for a first bandpass calibration
adopting the solution derived from 1921-293 and the self-calibration of the PKS~1830$-$211 data. 
At this step, the analysis of each 2-GHz spectral window revealed a residual bandpass, with amplitudes
roughly varying by an order of one percent over several 100~MHz. To remove this bandpass drift residuals
-- the final sensitivity being of about a few times $10^{-3}$ only --, data were then exported to FITS
format and subsequently processed with a customised ITT/IDL pipeline.

The pipeline is based on an iterative boxcar drift removal. Each iteration was designed to construct a model
spectrum (i.e., a spectrum with any slow variations in the bandpass removed) from the raw data following
several steps: 
\begin{itemize}
\item first, for each observed tuning, each polarization and each baseline, we flagged outliers in time by
assuming that the visibilities were constant on timescales longer than ten seconds;
\item a drift component was then computed and removed, in both time and frequency with different windows
widths, taking proper care of any flagged or missing data;
\item a first crude spectrum, after evaluating the time-averaged visibilities, was obtained and its residuals
were used to find frequency channels with large deviations from a random Gaussian distribution;
\item at this stage, we also examined the spectra for each antenna, to check for possible antenna-based problems;
\item the standard deviation of each baseline was then used to compute a weighted mean spectrum and the
uncertainty for each tuning and polarization;
\item finally, a Doppler track correction was applied according to the observatory velocities given in
Table~\ref{tab-journal}, obtained from the ATCA frequency calculator
\footnote{http://www.narrabri.atnf.csiro.au/observing/obstools/velo.html}.
All velocities in this paper refer to the Solar System barycenter rest-frame. We note that the Doppler correction
is nearly constant during the short observing intervals for individual tunings. Data were accordingly re-gridded
onto a common frequency scale using a one-dimensional (1D) drizzling algorithm (\citealt{fru02}). We tested different
frequency resolutions and chose a channel width of 0.5~MHz. A weighted mean was finally used to combine all spectra.
\end{itemize}
A model of the spectrum was derived by finding all frequency channels above a given threshold. This model
was then removed from the data, and the whole process was iterated on the residual datasets. 

We started with a large width for the bandpass residual filtering, corresponding roughly to 1~GHz, and a high
threshold of 10 $\sigma$ to derive the model. The pipeline was run until the variance of the resulting spectrum
was found to have reached a minimum. The model threshold was lowered iteratively in steps of two down to 4 $\sigma$
in order to correct for the negative corrections of the boxcar drift removal on strong lines. Following the same
scheme, the filtering width was successively divided by two until it reached a width of 16~MHz.

This whole procedure ensures that the second order bandpass is removed, without making any {\em a priori} assumptions
about the presence of strong absorption lines. The resulting spectrum is shown in Fig.\ref{fig-fullsurvey}.


\begin{table}[h] 
\caption{Journal of the observations.} \label{tab-journal}
\begin{center} \begin{tabular}{cccc}
\hline
Freq. & Time & $\Delta t$ & $V_{obs}$ \\
(GHz)   & (UT) & (h) & (km~s$^{-1}$) \\
\hline
\multicolumn{4}{c}{\em 2009 September 1} \\
\hline
31 \& 33   & 6:49  & 0.8 & $+25.71$ \\
37 \& 39   & 8:14  & 1.2 & $+25.84$ \\
44 \& 47.5 & 10:00 & 1.5 & $+26.03$ \\
42 \& 46.5 & 11:56 & 1.6 & $+26.23$ \\
35 \& 38   & 13:32 & 0.9 & $+26.36$ \\
\hline
\multicolumn{4}{c}{\em 2009 September 2} \\
\hline
32 \& 34   & 6:44  & 0.8 & $+25.95$ \\
41 \& 45   & 8:08  & 1.2 & $+26.07$ \\
46.5 \& 49 & 10:19 & 1.4 & $+26.30$ \\
43 \& 47   & 11:55 & 1.0 & $+26.47$ \\
36 \& 38.5 & 13:10 & 0.7 & $+26.57$ \\
\hline
\multicolumn{4}{c}{\em 2010 March 17} \\
\hline
37.25 \& 40  & 19:16 & 3.0 & $-29.44$ \\
46.1 \& 48.1 & 22:22 & 3.0 & $-29.16$ \\
\hline
\end{tabular}
\tablefoot{Frequencies indicate the center of each 2~GHz ATCA/CABB bands; time is given at the middle
of the time interval $\Delta t$ spent on the frequency tuning, and $V_{obs}$ is the observatory
velocity used to perform Doppler correction (with respect to the Solar System barycenter rest-frame).}
\end{center} \end{table}

\section{Results and analysis} \label{Results}

\begin{figure*}[h] \begin{center}
\includegraphics[height=\textwidth,angle=270]{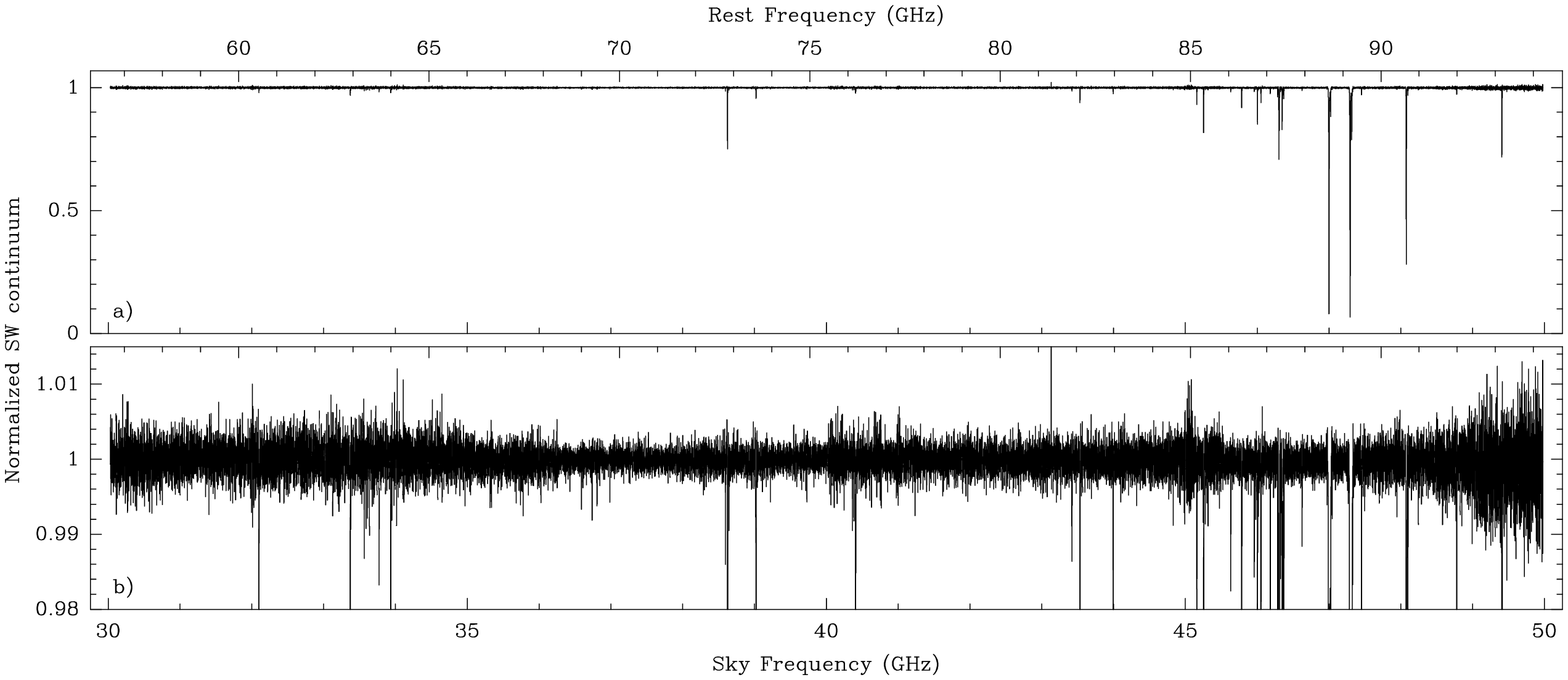}
\caption{a) Full 30--50 GHz spectral band observed toward PKS~1830$-$211 with the ATCA, normalized to
the continuum flux of the SW image of the quasar (38\% of the total flux). b) Same spectral interval,
but limited to a scale of a few percent of the SW continuum intensity, to illustrate the noise level
over the band. The rest-frame frequency axis is given on top of the figure.}
\label{overview}
\end{center} \end{figure*}

An overview of the 20-GHz-wide resulting spectrum is shown in Fig.\ref{overview}. The frequency coverage
corresponds, in the rest frame of the $z$=0.89 galaxy, to the range 57--94 GHz. The lower end of
this frequency interval is relatively unexplored because of poor atmospheric transmission caused by a forest
of fine-structure lines of O$_2$ around 60 GHz. The rms noise level is not uniform over the whole band,
and varies owing to the higher system temperatures at higher frequencies and the different integration times.
The full survey is presented in Fig.\ref{fig-fullsurvey}.

Only a handful of absorption lines reach an optical depth of 10\% (see Eq.\ref{eq-tau}) or more. They originate
in the SW absorption component and correspond, in order of decreasing optical depth, to: HCO$^+$, HCN, HNC,
C$_2$H, N$_2$H$^+$, H$_2$CO, c-C$_3$H$_2$, and H$^{13}$CO$^+$. All these species have been detected in previous
observations (\citealt{wik98,men99}). The weaker NE absorption at $V\sim-147$ km$^{-1}$ is detected only for
HCO$^+$, HCN, HNC, C$_2$H, c-C$_3$H$_2$, and H$_2$CO. 

\begin{figure}[h] \begin{center}
\includegraphics[width=8cm]{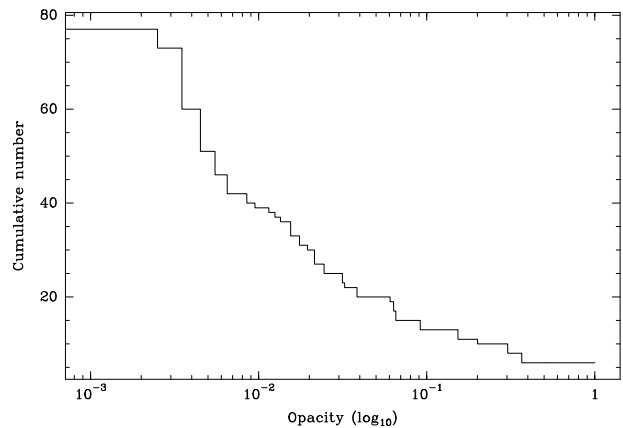}
\caption{Cumulative number of lines detected toward FG0.89SW with optical depths higher than a specified value.}
\label{fig-cumulative}
\end{center} \end{figure}

The major result of this survey, though, is the detection of a collection of weak absorption lines toward
the SW component, down to a few times 10$^{-3}$ of the continuum intensity, as illustrated in Fig.\ref{fig-cumulative},
where the cumulative number of detected lines is plotted as a function of opacity. These lines were identified
using the on-line Cologne Database for Molecular Spectroscopy\footnote{http://www.astro.uni-koeln.de/cdms}
(CDMS, \citealt{mul01}), and Jet Propulsion Laboratory Molecular Spectroscopy Database
\footnote{http://spec.jpl.nasa.gov/home.html} (JPL, \citealt{pic98}).
The corresponding transitions are listed in Table~\ref{tab-lines}.

All detected transitions (except the $K>1$ transitions of CH$_3$CN, see \S.\ref{ch3cn}) are ground state
or have low energy levels. Most molecules have high dipole moments, and the gas density is too moderate
for collisions to play an efficient role in the excitation. Rotational transitions are
therefore nearly in radiative equilibrium with the CMB, which has a temperature of $\sim$5~K at $z$=0.89.
Several species are detected in more than one transition, and a rotation diagram analysis indeed yields
rotation temperatures in agreement with this value. A low energy level is therefore a
good filter to identify transitions/species in the molecular databases, and this is why we believe
that our identifications are robust, even for species with a single detected transitions. The confusion
is also low, with a rough average of four clear lines per GHz over the 20-GHz-wide band.

The census of molecules detected to date in the $z$=0.89 galaxy located in front of PKS~1830$-$211
is given in Table~\ref{census}. It is quite remarkable that, despite its distance, this $z$=0.89
galaxy has now become the object with the largest number of detected molecules outside the Milky
Way. A large fraction of the total number of these species (28 out of 34) are detected in our survey,
among them 19 for the first time toward this source. As far as we know, this work also represents the
first extragalactic detection for SO$^+$, l-C$_3$H, l-C$_3$H$_2$, H$_2$CCN, H$_2$CCO, C$_4$H,
CH$_3$NH$_2$, and CH$_3$CHO. In addition, isotopologues of HCO$^+$, HCN, HNC, and SiO were also detected.
Elemental isotopic ratios of C, N, O, and Si are derived and discussed in \S\ref{section-isotopicratios}.

\begin{table*}[h]
\addtocounter{table}{+1}
\caption{Census of species detected toward FG0.89SW.} \label{census}
\begin{center} \begin{tabular}{ccccccc}
\hline

1 atom & 2 atoms & 3 atoms    & 4 atoms       & 5 atoms            & 6 atoms  & 7 atoms \\
 \hline
(H) $^{(d)}$ & (OH) $^{(d)}$   & (H$_2$O) $^{(h)}$  & (NH$_3$) $^{(gh)}$ & {\bf CH$_2$NH}       & {\bf CH$_3$OH} & {\bf \underline{CH$_3$NH$_2$}} \\
(C) $^{(i)}$ & (CO) $^{(bci)}$ & C$_2$H $^{(e)}$    & H$_2$CO $^{(ce)}$  & c-C$_3$H$_2$ $^{(e)}$ & {\bf CH$_3$CN} & {\bf CH$_3$C$_2$H} \\
            & (CS) $^{(af)}$  & HCN $^{(aef)}$ $^\triangle$     & {\bf \underline{l-C$_3$H}}      & {\bf \underline{l-C$_3$H$_2$}}   &                & {\bf \underline{CH$_3$CHO}} \\
            & SiO $^{(j)}$ $^\dagger$    & HNC $^{(aef)}$ $^\diamondsuit$     & {\bf HNCO}        & {\bf \underline{H$_2$CCN}}       &               & \\
            & {\bf NS}       & N$_2$H$^+$ $^{(a)}$ & {\bf H$_2$CS}     & {\bf \underline{H$_2$CCO}}       &               & \\
            & {\bf SO}       & HCO$^+$ $^{(aef)}$ $^\bigcirc$  &        & {\bf \underline{C$_4$H}}         &               & \\
            & {\bf \underline{SO$^+$}}   & {\bf HCO}          &                   & HC$_3$N $^{(ej)}$     &          & \\
            &                & {\bf HOC$^+$}      &                   &                      &          & \\
            &                & (H$_2$S) $^{(f)}$   &                   &                      &          & \\
            &                & {\bf C$_2$S}       &                   &                      &          & \\
\hline \end{tabular}
\tablefoot{In parenthesis, species detected in other studies;
bold face, new detections; underlined, first extragalactic detections.\\
$\bigcirc$ including H$^{13}$CO$^+$, HC$^{18}$O$^+$, and HC$^{17}$O$^+$;
$\triangle$ including H$^{13}$CN and HC$^{15}$N;
$\diamondsuit$ including HN$^{13}$C and H$^{15}$NC; 
$\dagger$ including $^{29}$SiO.\\
}
\tablebib{
(a) \cite{wik96b}; 
(b) \cite{ger97}; 
(c) \cite{wik98}; 
(d) \cite{che99}; 
(e) \cite{men99}; 
(f) \cite{mul06}, including isotopologues; 
(g) \cite{hen08}; 
(h) \cite{men08}; 
(i) \cite{bot09}; 
(j) \cite{hen09}. 
}
\end{center} \end{table*}

\subsection{Fitting procedure}

We fitted a Gaussian profile to all the transitions of a given species with a single velocity and a single
linewidth. The ortho and para forms, when relevant, were considered as the same species for the fit. The
relative intensities of all species with resolved or marginally resolved hyperfine structure were fixed,
as it was observed that, in the case of C$_2$H and HCO (for which the hyperfine structure is clearly resolved),
they follow the predicted ratios assuming local thermal equilibrium (LTE). The velocity and linewidth of
isotopic variants were fixed relative to the main isotopologue, except for HCO$^+$ and HCN, the lines of which
are likely saturated. For these, the corresponding isotopic variants were instead fixed relative to H$^{13}$CO$^+$
and H$^{13}$CN. For some species with data for the various transitions of low signal-to-noise ratio (S/N), we decided
to fix the velocity (e.g. l-C$_3$H$_2$) or the linewidth (e.g. SO$^+$) after initial free fits. Finally, the
fit for CH$_3$CN is degenerate because of the blending of the $K$=0 and $K$=1 transitions, and for this species,
we therefore chose to adjust a synthetic spectrum by eye. The fitting results are given in Table~\ref{tab-lines}.

\begin{table*}[h]
\caption{Molecular data for species detected toward the SW absorption.} \label{tab-moldata}
\begin{center} \begin{tabular}{lccccccccccc}
 
\hline
Molecule & $\mu$ & $Q_{5.14K}$ & $N_{line}$ & $V_0$ & $\Delta V$  & $T_{rot}$\tablefootmark{c} & $N_{COL}$\tablefootmark{c} & $N_{LTE}$\tablefootmark{d}  & Note \\
         & (Debye) &\tablefootmark{a} & \tablefootmark{b} & (km~s$^{-1}$)  & (km~s$^{-1}$) &  (K)     & ($10^{12}$ cm$^{-2}$) & ($10^{12}$ cm$^{-2}$) & \\
\hline
              C$_2$H& 0.77&  2.81& 6&$  -0.45\ (  0.03)$&$  20.9\ (   0.1)$&--&--&$  1248.5 \ (     1.4$)&            hfs,CDMS\\
                 HCN& 2.99&  2.78& 1&$  -2.83\ (  0.01)$&$  28.9\ (   0.0)$&--&--&$   304.1 \ (     0.5$)&                CDMS\\
          H$^{13}$CN& 2.99&  2.84& 1&$  -2.27\ (  0.14)$&$  21.4\ (   0.3)$&--&--&$     8.4 \ (     0.1$)&                CDMS\\
          HC$^{15}$N& 2.99&  2.85& 1&--&--&--&--&$     1.6 \ (     0.1$)&                CDMS\\
                 HNC& 3.05&  2.73& 1&$  -1.75\ (  0.02)$&$  20.8\ (   0.1)$&--&--&$   101.8 \ (     0.3$)&                CDMS\\
          HN$^{13}$C& 3.05&  2.82& 1&--&--&--&--&$     2.2 \ (     0.1$)&                CDMS\\
          H$^{15}$NC& 3.05&  2.77& 1&--&--&--&--&$     0.3 \ (     0.1$)&                 JPL\\
          N$_2$H$^+$& 3.40&  2.66& 1&$  -3.29\ (  0.12)$&$  21.0\ (   0.3)$&--&--&$    23.3 \ (     0.3$)&                CDMS\\
             HCO$^+$& 3.90&  2.77& 1&$  -2.43\ (  0.02)$&$  32.4\ (   0.0)$&--&--&$   175.1 \ (     0.3$)&                CDMS\\
      H$^{13}$CO$^+$& 3.90&  2.83& 1&$  -1.83\ (  0.07)$&$  18.4\ (   0.2)$&--&--&$     8.3 \ (     0.1$)&                CDMS\\
      HC$^{18}$O$^+$& 3.90&  2.88& 1&--&--&--&--&$     3.6 \ (     0.1$)&                CDMS\\
      HC$^{17}$O$^+$& 3.90&  2.82& 1&--&--&--&--&$     0.2 \ (     0.0$)&                CDMS\\
             HOC$^+$& 2.77&  2.76& 1&$  -0.78\ (  0.39)$&$  20.5\ (   0.9)$&--&--&$     3.2 \ (     0.1$)&                CDMS\\
                 HCO& 1.36&  2.83& 4&$   1.44\ (  0.59)$&$  18.1\ (   1.4)$&--&--&$    13.0 \ (     1.8$)&             hfs,JPL\\
            CH$_2$NH& 1.53& 14.63& 1&$  -0.71\ (  0.36)$&$  13.8\ (   0.8)$&--&--&$    14.9 \ (     0.8$)&                 JPL\\
          H$_2$CO(p)& 2.33&  3.30& 1&$  -1.00\ (  0.04)$&$  20.3\ (   0.1)$&--&--&$    59.8 \ (     0.2$)&                CDMS\\
     CH$_3$NH$_2$(o)& 1.26&208.60& 2&$  -6.19\ (  1.64)$&$  20.6\ (   3.9 )$&   4.0\ $_{-   3.0}^{+  11.0}$&     6.0\ $_{-     4.0}^{+    18.0}$&$    15.0 \ (     3.1$)&                 JPL\\
     CH$_3$NH$_2$(p)& 1.26& 69.84& 1&--&--&--&--&$    17.1 \ (     3.9$)&                 JPL\\
            CH$_3$OH& 1.44&  6.97& 1&$  -5.26\ (  0.51)$&$  21.2\ (   1.2)$&--&--&$   170.0 \ (     8.3$)&                CDMS\\
            l-C$_3$H& 3.55& 43.53& 4&$   0.18\ (  0.50)$&$  17.5\ (   1.0)$&--&--&$     7.1 \ (     0.1$)&            hfs,CDMS\\
     c-C$_3$H$_2$(o)& 3.43& 22.57& 2&$  -0.94\ (  0.07)$&$  18.7\ (   0.2 )$&   5.6\ $_{-   0.4}^{+   0.4}$&    32.0\ $_{-     4.0}^{+     4.0}$&$    39.1 \ (     0.3$)&                 JPL\\
     c-C$_3$H$_2$(p)& 3.43&  7.53& 2&--&--&   5.4\ $_{-   1.0}^{+   1.0}$&    10.5\ $_{-     3.0}^{+     3.5}$&$    13.8 \ (     0.2$)&                 JPL\\
     l-C$_3$H$_2$(o)& 4.10&  4.30& 3&$  -1.93\ (  0.00)$&$  21.1\ (   2.6 )$&   5.3\ $_{-   3.0}^{+  11.5}$&     1.0\ $_{-     0.5}^{+     1.5}$&$     1.5 \ (     0.2$)&                CDMS\\
     l-C$_3$H$_2$(p)& 4.10& 10.64& 1&--&--&--&--&$     0.6 \ (     0.2$)&                CDMS\\
         H$_2$CCN(o)& 3.50&197.52& 2&$  -1.83\ (  2.03)$&$  17.9\ (   4.5 )$&   3.0\ $_{-   1.2}^{+   3.1}$&     1.0\ $_{-     0.3}^{+     0.9}$&$     2.2 \ (     0.3$)&            hfs,CDMS\\
           CH$_3$CCH& 0.78& 85.55& 1&$   0.59\ (  1.54)$&$  15.6\ (   4.0)$&--&--&$    60.0 \ (     0.0$)&                CDMS\\
            CH$_3$CN& 3.92& 80.51& 2&$  -2.70\ (  0.00)$&$  21.6\ (   0.0)$&--&--&$    13.0 \ (     0.0$)&                CDMS\\
         H$_2$CCO(o)& 1.42&  4.71& 4&$  -1.62\ (  0.99)$&$  21.2\ (   2.3 )$&   2.6\ $_{-   0.8}^{+   1.1}$&     5.0\ $_{-     1.5}^{+     2.0}$&$    13.2 \ (     1.1$)&                CDMS\\
                HNCO& 1.58& 10.09& 2&$  -5.12\ (  0.88)$&$  22.0\ (   2.1)$&--&--&$     6.2 \ (     0.6$)&                CDMS\\
                 SiO& 3.10&  5.28& 1&$  -2.29\ (  0.18)$&$  20.6\ (   0.4)$&--&--&$     7.5 \ (     0.1$)&                CDMS\\
          $^{29}$SiO& 3.10&  5.34& 1&--&--&--&--&$     0.7 \ (     0.2$)&                CDMS\\
           CH$_3$CHO& 2.42& 55.30& 9&$  -0.80\ (  0.73)$&$  19.3\ (   1.7 )$&   7.5\ $_{-   3.5}^{+  12.0}$&    15.0\ $_{-     9.0}^{+    30.0}$&$    12.4 \ (     0.9$)&                 JPL\\
                  NS& 1.81& 26.85&10&$  -2.61\ (  0.78)$&$  20.9\ (   1.8)$&--&--&$    11.5 \ (     0.6$)&             hfs,JPL\\
          H$_2$CS(o)& 1.65&  2.58& 2&$  -1.60\ (  0.79)$&$  14.1\ (   1.9)$&--&--&$     5.7 \ (     0.8$)&                CDMS\\
          H$_2$CS(p)& 1.65&  6.58& 1&--&--&--&--&$     1.9 \ (     0.3$)&                CDMS\\
                  SO& 1.54&  7.64& 2&$  -1.01\ (  0.30)$&$  19.7\ (   0.7 )$&   5.4\ $_{-   1.4}^{+   1.4}$&    21.0\ $_{-    10.0}^{+    10.0}$&$    25.6 \ (     0.8$)&                CDMS\\
              SO$^+$& 2.30& 11.06& 2&$   1.54\ (  2.00)$&$  20.0\ (   0.0)$&--&--&$     1.7 \ (     0.3$)&                 JPL\\
              C$_4$H& 0.87& 91.36& 8&$  -1.05\ (  0.59)$&$  18.5\ (   1.4 )$&   8.0\ $_{-   3.5}^{+   6.5}$&   110.0\ $_{-    30.0}^{+    90.0}$&$   131.3 \ (     7.8$)&                CDMS\\
             HC$_3$N& 3.73& 23.88& 4&$  -3.90\ (  0.46)$&$  21.4\ (   0.9 )$&   6.3\ $_{-   1.3}^{+   1.3}$&     8.3\ $_{-     0.6}^{+     0.6}$&$    11.2 \ (     0.4$)&                CDMS\\
              C$_2$S& 2.88& 24.71& 3&$  -1.35\ (  0.83)$&$  18.2\ (   2.0 )$&   5.7\ $_{-   3.0}^{+  12.0}$&     3.2\ $_{-     1.3}^{+     4.6}$&$     4.8 \ (     0.4$)&                CDMS\\
\hline
\end{tabular}
\tablefoot{
\tablefoottext{a}{For species with ortho/para symmetry, the two forms were considered as
distinct species and the partition function Q was thus calculated accordingly, with
the weight factors $g$=3 for ortho form, and $g$=1 for para form.
Hyperfine structure was taken into account when necessary.}\\
\tablefoottext{b}{Number of lines detected or resolved (in the case of hyperfine structure).}\\
\tablefoottext{c}{Rotation temperature and column density obtained from Monte Carlo simulations of rotation diagrams. Uncertainties
correspond to a 95\% confidence interval.}\\
\tablefoottext{d}{Column density, assuming a rotation temperature of 5.14~K and LTE.}\\
\tablefoottext{e}{Possibly underestimated due to the large opacity of the line.}\\
}
\end{center} \end{table*}

\begin{table*}[h]
\caption{Line parameters and fitting results for the NE absorption component.} \label{tab-NE}
\begin{center} \begin{tabular}{lcccccc}
\hline
Species  & $V_0$ & $\Delta V$ & $\int \tau $d$V$ & $N_{LTE}$ & $N_{LTE}^{\rm SW}$/$N_{LTE}^{\rm NE}$ \\
         & (km~s$^{-1}$) & (km~s$^{-1}$)   & (10$^{-3}$ km~s$^{-1}$)    & ($10^{12}$ cm$^{-2}$) & \\
\hline
             HCO$^+$&$   -148.1\ (    0.1$)&$ 21.5\ (  0.1)$&$2897.0\ (  15.0)$&$   7.5\ (   0.0)$&$ 23.4\ (  0.1)$\\
                 HCN&$   -147.5\ (    0.1$)&$ 22.2\ (  0.2)$&$1656.0\ (  14.0)$&$   7.3\ (   0.1)$&$ 41.5\ (  0.4)$\\
                 HNC&$   -147.0\ (    0.5$)&$ 15.0\ (  1.0)$&$ 284.0\ (  14.0)$&$   1.2\ (   0.1)$&$ 87.2\ (  4.3)$\\
              C$_2$H&$   -148.1\ (    0.5$)&$ 15.0\ (  0.0)$&$ 429.0\ (   9.0)$&$  29.2\ (   0.6)$&$ 42.7\ (  0.9)$\\
     c-C$_3$H$_2$(o)&$   -147.5\ (    1.4$)&$ 16.0\ (  3.4)$&$  84.0\ (  18.0)$&$   0.8\ (   0.2)$&$ 47.9\ ( 10.3)$\\
     c-C$_3$H$_2$(p)&$   -147.5\ (    1.4$)&$ 16.0\ (  3.4)$&$  38.0\ (  11.0)$&$   0.4\ (   0.1)$&$ 31.9\ (  9.2)$\\
          H$_2$CO(p)&$   -146.7\ (    0.7$)&$ 15.4\ (  1.7)$&$ 107.0\ (  10.0)$&$   1.1\ (   0.1)$&$ 56.6\ (  5.3)$\\
          N$_2$H$^+$&$   -147.0\ (    0.0$)&$ 20.0\ (  0.0)$&$ 105.0\ (   0.0)$&$  -0.3\ (   0.0)$&$-69.9\ (  0.0)$\\
      H$^{13}$CO$^+$&$   -147.0\ (    0.0$)&$ 20.0\ (  0.0)$&$  39.0\ (   0.0)$&$  -0.1\ (   0.0)$&$-78.9\ (  0.0)$\\
          H$^{13}$CN&$   -147.0\ (    0.0$)&$ 20.0\ (  0.0)$&$  38.1\ (   0.0)$&$  -0.2\ (   0.0)$&$-47.8\ (  0.0)$\\
\hline \end{tabular} \end{center} \tablefoot{Negative values (except for velocities) represent either the (3 $\sigma$) upper (for $N_{LTE}$) or lower (for [SW]/[NE]) limits.} \end{table*}

\begin{table*}[h]
\caption{Upper limits (3$\sigma$) for non-detections of some other interstellar species toward FG0.89SW.} \label{tab-uplim}
\begin{center} \begin{tabular}{lccccccccc}
\hline
Species  & Rest Freq. & $\mu$ & $Q_{5.14K}$ & $S_{ul}$ & $E_l/k_B$ & $\int \tau $d$V$ & $N_{LTE}$ & [X]/[H$_2$] \\
         & (MHz)      & (D)   &             &          & (K)       & (10$^{-3}$ km~s$^{-1}$)    & ($10^{12}$ cm$^{-2}$)    & $10^{-11}$    \\
\hline
             DCO$^+$~                                                  &72039.314& 3.90&  3.3& 1.00&  0.0& $<$ 103.5& $<$  0.1& $<$   0.62\\
                 DCN~                                                  &72414.696& 2.99&  3.3& 1.00&  0.0& $<$ 126.0& $<$  0.3& $<$   1.27\\
               O$_2$~                                                  &58446.587&--&  5.0& 0.00& 23.6& $<$ 205.2& $<$     265940.& $<$    1329702.\\
              C$_3$S~                                                  &57806.705& 3.70& 37.4&10.02& 12.5& $<$ 225.0& $<$  4.5& $<$  22.35\\
             HC$_5$N~                                                  &58577.354& 4.33& 80.8&22.00& 29.5& $<$ 146.7& $<$ 56.6& $<$ 283.08\\
              SO$_2$~                                                  &69575.928& 1.63& 13.6& 1.00&  0.0& $<$ 109.8& $<$  3.1& $<$  15.70\\
                 OCS~                                                  &72976.776& 0.72& 17.9& 6.00&  8.8& $<$ 140.4& $<$ 24.3& $<$ 121.31\\
           $^{34}$SO~                                                  &62074.047& 1.54&  7.7& 1.93&  1.4& $<$ 142.2& $<$  1.9& $<$   9.59\\
         CH$_2$D$^+$~                                                  &67273.575& 0.33&  0.4& 4.50& 18.6& $<$ 139.5& $<$ 23.0& $<$ 115.03\\
     H$_2^{13}$CO(p)~                                                  &71024.788& 2.33&  3.4& 1.00&  0.0& $<$ 100.8& $<$  0.3& $<$   1.73\\
            HCNH$^+$~                                                  &74111.305& 0.29&  3.2& 1.00&  0.0& $<$ 123.3& $<$ 25.5& $<$ 127.34\\
      $^{15}$NNH$^+$~                                                  &90263.840& 3.40&  2.7& 1.00&  0.0& $<$ 225.0& $<$  0.2& $<$   1.25\\
      N$^{15}$NH$^+$~                                                  &91205.696& 3.40&  2.7& 1.00&  0.0& $<$ 246.6& $<$  0.3& $<$   1.35\\
          $^{13}$CCH~                                                  &84119.331& 0.77& 23.1& 2.00&  0.0& $<$ 225.0& $<$ 21.5& $<$ 107.42\\
          C$^{13}$CH~                                                  &85229.332& 0.77& 23.0& 2.00&  0.0& $<$ 303.3& $<$ 28.6& $<$ 142.96\\
        C$_2$H$_3$CN~                                                  &56786.934& 3.82&178.7&17.95&  6.8& $<$1201.5& $<$ 20.0& $<$  99.88\\
        HC$_3$NH$^+$~                                                  &60605.339& 1.61& 73.2&20.70&  8.7& $<$ 148.5& $<$  6.8& $<$  34.03\\
              C$_2$O~                                                  &69069.473& 1.31& 28.5& 3.84&  3.3& $<$ 109.8& $<$  5.1& $<$  25.35\\
\hline \end{tabular} \end{center} \end{table*}

\subsection{Time variations} \label{section-timvar}

A monitoring of the HCO$^+$ $J$=2-1 absorption profile between 1995 and 2008 revealed significant time
variations (\citealt{mul08}), in particular with the quasi-disappearance of the $-147$ km\,s$^{-1}$
absorption component between 2006 to 2007. In the HCO$^+$ $J$=2-1 spectrum acquired by
Muller \& Gu\'elin in October 2008 with the Plateau de Bure interferometer, this component had returned
with an absorption of $\sim$15\% of the total continuum intensity. Variations were found to be
correlated between the $-147$ km\,s$^{-1}$ component and the blue wings of the 0 km\,s$^{-1}$ absorption,
hence should originate from morphological changes in the background quasar.

From Very Long Baseline Interferometry continuum measurements, the relative distance between the NE and
SW lensed images of the quasar was found to vary by as much as 200 micro-arcseconds within a few months
(\citealt{jin03}). \cite{nai05} proposed that this apparent motion is caused by the sporadic ejection
of plasmons along a helical jet by the quasar. Consequently, the continuum illumination scans different
pencil beams through the foreground absorbing clouds, where the apparent displacement of the quasar
lensed images corresponds to a projected distance of about a parcsec, i.e. a scale comparable to the
size of Galactic molecular-cloud cores.

The relative differences (normalized to the averaged spectrum) for the spectral regions covered in both
runs (September 2009 and March 2010) are shown together with the averaged spectrum in
Fig.\ref{fig-fullsurvey}. They are at most $\sim$5\% for the HCO$^+$, HCN, and HNC lines, and barely
visible for all other lines. Given the small relative changes, data from both observing sessions were
combined to improve the S/N of the data.

\subsection{Background continuum illumination} \label{back-illumin}

Absorption intensities (I$_\nu$), measured from the continuum level, were converted to optical depths
($\tau_\nu$) according to
\begin{equation}
 \tau_\nu = - \ln{ \left ( 1-\frac{I_\nu}{f_c I_{bg}} \right ), }
\label{eq-tau}
\end{equation}
\noindent where $I_{bg}$ is the intensity of the background continuum source, and $f_{c}$ the source
covering factor. At radio wavelengths, the continuum emission of the quasar PKS~1830$-$211, lensed
by the $z$=0.89 intervening galaxy, is dominated by two bright and compact components (to the NE
and SW), separated by 1'' and included in an overall fainter Einstein ring. The flux ratio of the
NE to SW compact components is $\sim$1.5--1.6 at cm wavelengths as measured by \cite{sub90}. The
remaining emission (Einstein ring and fainter other components) contributes only to a few percent
of the total continuum intensity.

Molecular absorption in the $z$=0.89 galaxy is observed toward both the SW and NE continuum images.
The HCO$^+$ $J$=2-1 absorption toward the SW image shows a flat-bottom profile, indicating saturation.
Assuming that, at this wavelength, all the continuum emission comes from the NE and SW images of the
quasar, it is then possible to derive the relative flux NE/SW of both images from the level of this
saturation, without actually resolving them. The NE/SW ratio was measured to be about 1.7 with a
dispersion of 0.3 over several observations between 1995 and 2007 (\citealt{mul08}), i.e. roughly
similar to the flux ratio directly measured at cm wavelengths. Some of the fluctuations might be
due to morphological changes in the quasar (plasmon burst, microlensing) and the time delay between
the two line of sights.

Our 7 mm survey is limited in terms of spectral resolution, and the flat-bottom part of the
HCO$^+$/HCN $J$=1-0 lines (supposing that they are also saturated) is smooth, preventing the same
analysis. Nevertheless, as the HCO$^+$ $1-0$ absorption reaches an apparent depth of $\sim$35\%
and 8\% toward the SW and NE components, relative to the total continuum level, we can certainly
state that $0.35 < I_{bg}({\rm SW}) < 0.92$ and $0.08 < I_{bg}({\rm NE}) < 0.65$. If,
in addition, we impose the ratio $R=$NE/SW, we get the stronger constraints of
$0.35 < I_{bg}({\rm SW}) < \frac{0.65}{R}$ and $0.35\times R < I_{bg}({\rm NE}) < 0.65$, which gives
$I_{bg}$(NE)=$0.61 \pm 0.05$ and $I_{bg}$(SW)=$0.38 \pm 0.03$ with $R$=1.6. We adopt these values.
The rough uncertainty in the illuminating background continuum for both NE and SW components is
therefore less than 10\%. The residual contribution from components other than the NE and SW images
is {\em de facto} less than $[1-0.35\times(1+R)]$, that is 9\% with $R$=1.6.

This is valid, of course, if no changes happened in the quasar within the time corresponding to the
time delay between the two lines of sight ($\sim$24 days) before our observations. Unfortunately,
we do not have flux monitoring data corresponding to these epochs. Nevertheless, that we do not see
significant variations in the absorption line profiles between the observations of September 2009
and March 2010 is reassuring.

We can now investigate how this uncertainty in the background continuum illumination propagates to the
derivation of opacities, by means of Eq.\ref{eq-tau}. In the case of weak lines ($I<<I_{bg}$), we get
$\frac{\Delta \tau}{\tau} = \frac{\Delta I_{bg}}{I_{bg}}$, and our opacities are therefore estimated
more accurately than 10\%. For the other lines, we get $\frac{\Delta \tau}{\tau} = \frac{\Delta I_{bg}}{I_{bg}}
\times F(I/I_{bg})$, where the function $F(x) = x / ((x-1)\ln{(1-x)})$, and tends to infinity for
$x \rightarrow 1$, although the rise is slow, as $F(0.9)$ (that is $I/I_{bg}= 90\%$) is still $<4$.
For example, the corresponding uncertainty in HNC opacity (the third strongest absorption line in
our survey, after HCO$^+$ and HCN) is less than 20\%.

Finally, we assume here (as in the rest of the paper) a source covering factor $f_{c}$ of unity for all absorption 
components. It is however possible that $f_{c}<1$ and changes with frequency, as the continuum emission
probably becomes more extended at lower frequencies (see also the discussion by \citealt{hen09}).
As a result, the optical depths derived from Eq.\ref{eq-tau} are, strictly speaking, lower limits.
We note however that the HCO$^+$ and HCN $J$=2-1 lines toward the SW absorption are optically thick and
saturated.
The source covering factor for this component
is thus certainly close to unity. Toward the NE absorption, the situation is not so clear. The formal
observational constraints from the HCO$^+$ 1-0 line are $\tau>0.14$ and $f_{c}>0.13$. While the non-detection
of H$^{13}$CO$^+$ (with HCO$^+$/H$^{13}$CO$^+$$>$71) suggests that the line is likely optically thin, the
source covering factor is still only poorly constrained.

\subsection{LTE analysis and rotation diagrams}

For optically thin lines, and assuming a Boltzmann distribution characterized by a rotation temperature $T_{rot}$,
column densities $N_{col}$ can be derived as
\begin{equation}
N_{LTE} = \frac{3h}{8\pi^3\mu^2S_{ul}}\frac{Q(T_{rot})\exp{(\frac{E_l}{k_BT_{rot}})}}{[1-\exp{(\frac{-h\nu}{k_{B}T_{rot}})}]} \int \tau dV, 
\label{eq-ncol}
\end{equation}
\noindent where $Q(T_{rot})$ is the partition function, $E_{l}$ the energy of the lower level with respect
to ground state, $\mu$ the dipole moment, $\nu$ the frequency, and $S_{ul}$ the line strength.
Under these conditions, the column density of a detected molecule can be calculated directly with the
knowledge of only one physical parameter: the rotation temperature $T_{rot}$.

Since some species are detected in several transitions, an excitation analysis can be done.
Under the Rayleigh-Jeans (RJ) approximation ($h\nu<<k_{B}T$), Eq.\ref{eq-ncol} can be re-arranged as
\begin{equation}
\log{\left (\frac{3k_{B}\int \tau dV}{8\pi^3 \nu \mu^2S_{ul}} \right )} = \log{\left (\frac{N_{col}}{Q(T_{rot}) T_{rot}}\right )} - \frac{E_{l} \log{(e)}}{k_{B} T_{rot}}.
\label{eq-trot}
\end{equation}
\noindent Plotting the quantity on the left side of this equation versus the energy of the lower level
for each observed transition then allows the rotation temperature $T_{rot}$ to be measured as the inverse
of the slope of a linear fit to the data points in the rotation diagram. Column densities were also
derived, following Eq.\ref{eq-trot}, from the intercept of the y-axis in the rotation diagram. Because
the uncertainties in the data points in the rotation diagrams do not follow a normal distribution, the
rotation temperatures, column densities, and their corresponding uncertainties were estimated using
Monte Carlo simulations. We note that the RJ approximation is still valid for $T>4.5$ K at 94 GHz, the
highest rest-frame frequency reached in our survey.

\begin{figure*}[h] \begin{center}
\includegraphics[width=\textwidth]{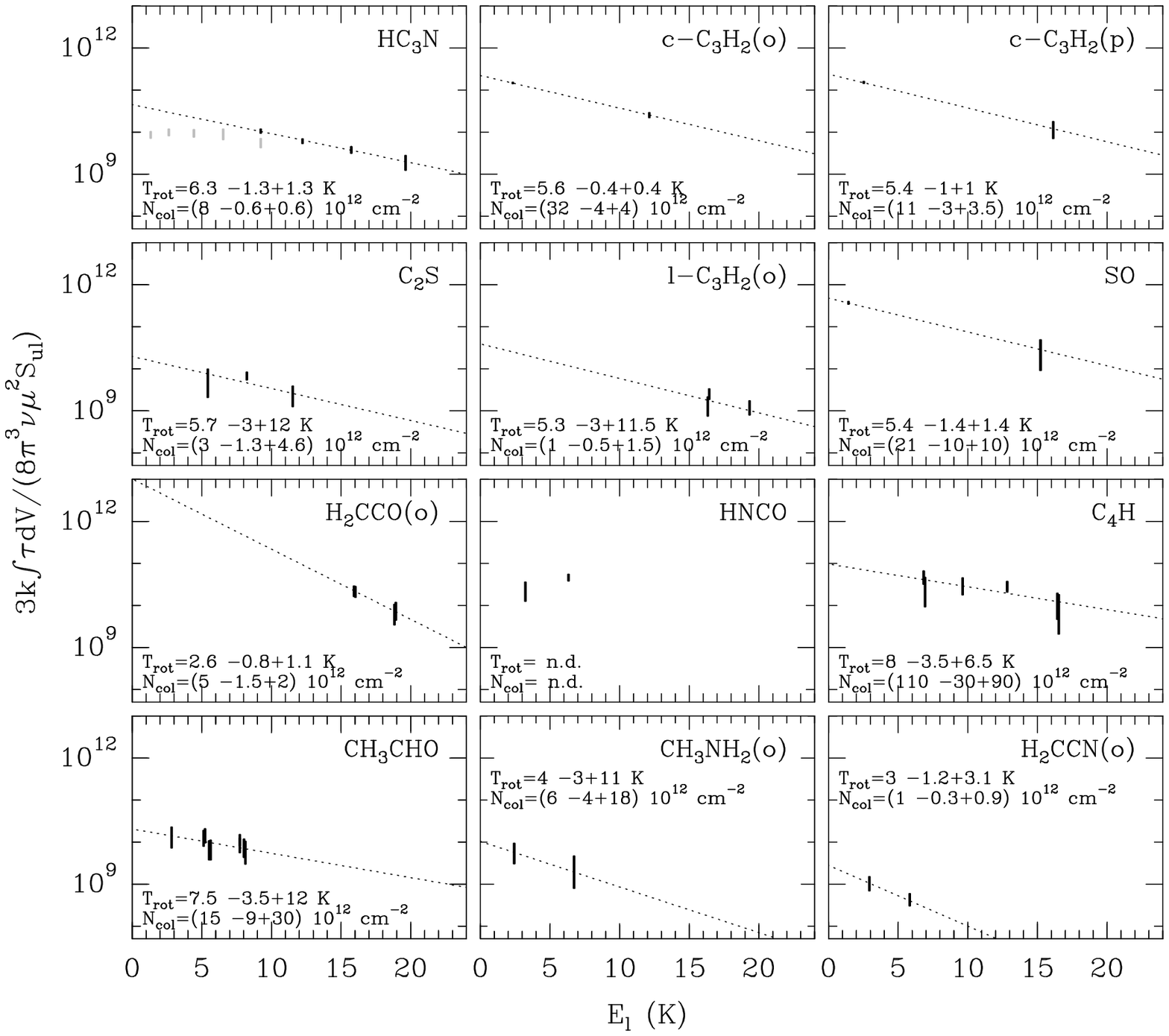}
\caption{Rotation diagrams. Uncertainties in the rotation temperatures and column densities were obtained from 
Monte Carlo simulations, and are given for a 95\% confidence level. Data points for additional transitions of
HC$_3$N observed by \cite{hen09} are indicated in light grey (see \ref{appendix-HC3N}).}
\label{fig-rotdiag}
\end{center} \end{figure*}

\begin{figure}[h] \begin{center}
\includegraphics[width=8cm]{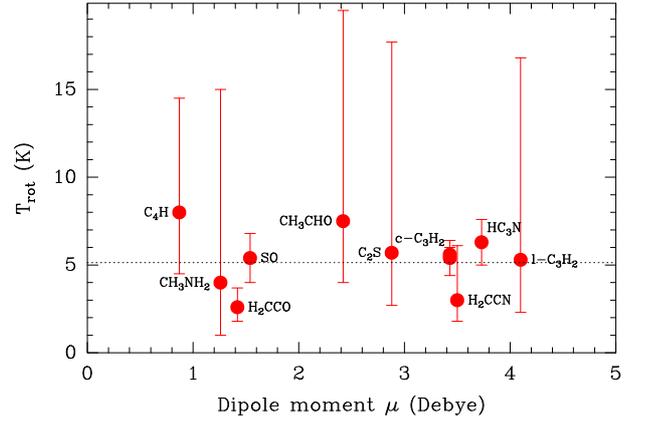}
\caption{Rotation temperatures as a function of molecular dipole moment.
The dashed line indicates the value $T_{CMB}$=5.14 K at $z$=0.89, predicted from standard cosmology.
Error bars indicate the 95\% confidence interval.}
\label{fig-Trotvsmu}
\end{center} \end{figure}

The rotation diagrams for molecules detected toward FG0.89SW in several transitions are presented in
Fig.\ref{fig-rotdiag}, and the resulting measurements of rotation temperatures and column densities are
listed in Table~\ref{tab-moldata}. These rotation temperatures are plotted in Fig.\ref{fig-Trotvsmu} as
a function of the molecular dipole moment. The results for column densities toward FG0.89NE are given
in Table~\ref{tab-NE}.

Most of these species have high dipole moments, and consequently high critical densities ($>$10$^5$~cm$^{-3}$).
For low density gas ($\sim$10$^3$ cm$^{-3}$), the population of the different energy levels is not efficiently
thermalized by collisions, but is coupled with the ambient radiation field, which, in the absence of any other
radiative excitation (UV or IR pumping), consists of CMB photons. In this case, rotation temperatures then nearly
equal the CMB temperature, which is expected to be $T_{CMB}$=5.14~K at $z$=0.89, assuming a scaling proportional
to $(1+z)$. The derived rotation temperatures are consistent with this value. A thorough excitation analysis is
beyond the scope of this paper, as it would require a careful treatment of collisional rates and partners (H$_2$
and e$^-$) for each molecule. This will be the subject of a future paper, in which we will also study the
excitation in FG0.89NE with new multi-transitions observations, in order to derive an accurate and robust
measurement of the CMB temperature at $z$=0.89 through the two independent lines of sight.

As a first step to interpreting the observations, we produced a LTE synthetic spectrum resulting from the
combination of absorption lines from the molecular species listed in Table~\ref{tab-moldata}. We assume a
common rotation temperature $T_{rot}$=$T_{CMB}$=5.14~K for all species and for the sake of simplicity, the
absorption profile is set to be Gaussian with a FWHM of 20 km\,s$^{-1}$ centered at $V$=0~km\,s$^{-1}$ for
the SW absorption (some lines are slightly offset from these values, see Table~\ref{tab-moldata} and
\S\ref{section-velo}) and at $V$=$-147$~km\,s$^{-1}$ for the few species for which NE absorption is detected.
The partition functions were calculated for $T$=5.14~K by performing a direct sum over all energy levels.
Their values are given in Table~\ref{tab-moldata}. Following Eq.\ref{eq-ncol}, the only free parameter is
then the column density of each species. A kinetic temperature of 50~K was adopted to take into account
the $K$-ladder population of the symmetric-top molecule CH$_3$CN (see \S\ref{ch3cn}).

The resulting LTE spectrum, obtained with the LTE column densities listed in Table~\ref{tab-moldata}, 
is overlaid on top of the observed spectrum in Fig.\ref{fig-fullsurvey}.
Despite the simplicity of the model, it reproduces relatively well the observations.

In addition, we list in Table~\ref{tab-uplim} the upper limits to the column densities and abundances obtained
for some interstellar species of interest that have low-energy transitions within the frequency coverage of
our survey. The upper limits to the integrated intensities were calculated as
$3\sigma_{\tau}\sqrt{\delta V \Delta V_{1/2}}$, where $\sigma_{\tau}$ is the optical depth (Eq.\ref{eq-tau})
noise level, $\delta V$ is the velocity resolution, and $\Delta V_{1/2}$ the linewidth at half-maximum, which
we fixed to 20~km\,s$^{-1}$. The upper limits to the column densities were calculated from Eq.\ref{eq-ncol},
again assuming LTE and $T_{rot}$=5.14~K.

\subsection{Additional velocity components} \label{section-newvelocomp}

\begin{figure}[h] \begin{center}
\includegraphics[width=8cm]{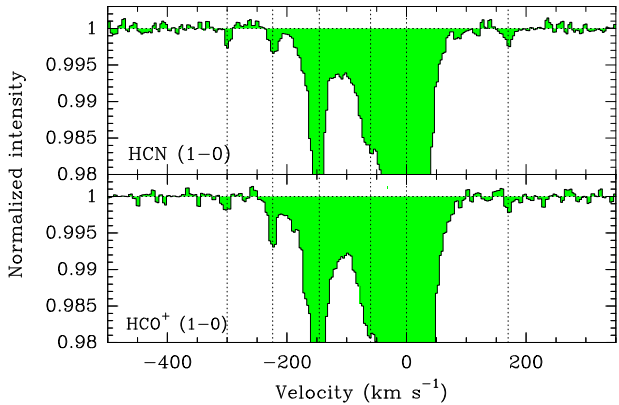}
\caption{Spectra of the HCO$^+$ and HCN 1-0 lines, showing the absorption components at 0 and $-147$~km\,s$^{-1}$
(previously known, and arising toward the SW and NE images of the quasar, respectively), as well as the additional
absorption components at velocities $-300$, $-224$, $-60$, and +170~km\,s$^{-1}$.}
\label{fig-newvelocomp}
\end{center} \end{figure}

Previous observations have shown that the molecular absorption profile consists of two main velocity components,
separated by 147~km\,s$^{-1}$ and located in front of the SW and NE images of the quasar (see Introduction).
However, this picture should now be updated with the discovery of additional absorption components.
Close to the redshifted frequencies of both the HCN and HCO$^+$ 1-0 lines (redshifted to $\sim$47
and 47.3~GHz respectively) in Fig.\ref{fig-fullsurvey}, several additional weak lines can be seen, which we failed
to identify with other species. Those lines clearly arise at the same velocities for both HCN and
HCO$^+$ 1-0 transitions, as shown in Fig.\ref{fig-newvelocomp}. They therefore correspond to 
additional velocity components of HCN and HCO$^+$ 1-0 in the $z$=0.89 galaxy. The new
components are located at $-300$, $-224$, $-60$, and +170~km\,s$^{-1}$ (in addition to the previously known ones at 0 and
$-147$~km\,s$^{-1}$).
They probably arise in front of secondary continuum peaks along the Einstein ring (see for example Fig.3
by \citealt{che99}). It would be interesting to determine their locations to constrain the
kinematics of the galaxy, as well as their background continuum illumination to derive the
associated opacities and column densities of gas along different sightlines through the disk. This is
unfortunately impossible with the current data, which are limited in terms of angular resolution and sensitivity.
Interestingly, the $-300$ and +170~km\,s$^{-1}$ components have velocities beyond the range covered by HI absorption
(\citealt{che99,koo05}), and are nearly symmetrical with respect to the central velocity ($\sim$$-$80~km\,s$^{-1}$ from the
velocity of the SW absorption). These two components may arise from rapidly rotating gas near the center of the galaxy,
e.g., in a circumnuclear ring.

Finally, we emphasize that we were unable to identify any lines originating from the second intervening galaxy
at $z$=0.1926 (\citealt{lov96}), 
neither from the host galaxy of the quasar at $z$=2.5 (\citealt{lid99}), nor from Galactic absorption.

\section{Discussion} \label{Discussion}

The advantages of our survey are manifold. First, the 4-mm rest-frame window (redshifted to 7 mm at
$z$=0.89), includes numerous ground state and low energy level transitions of a large range of
intermediate-size interstellar molecules, and is therefore particularly interesting for an absorption
study. Lighter molecules, e.g. hydrides, on the one hand, and heavier and more complex molecules (with
$\geq 6$ atoms) on the other hand, generally have these transitions at higher (mm/submm window) and
lower ($\sim$cm wavelengths) frequencies, respectively. The achieved sensitivity, of a fraction of a
percent relative to the continuum intensity, and the large spectral coverage (20 GHz band) then allows
us to detect many weak lines of rare molecules (i.e. with abundances of a few 10$^{-11}$ relative to
H$_2$) and in several cases, to build their rotation diagrams. For both runs (September 2009 and March
2010), observations were carried out within one or two days, i.e. a timescale shorter than the time
delay of the lens ($\sim$24 days, \citealt{lov98,wik99}) and much smaller than the timescale for
variations in the absorption profiles, of the order of several months (\citealt{mul08}). The comparison
of lines in the frequency overlap between the two datasets shows that the relative variations between
September 2009 and March 2010 are smaller than $\sim$5\% for the strong HCO$^+$/HCN 1-0 lines and yet
smaller for other lines. Time variations are therefore negligible, and moreover, would affect only the
few lines in the small frequency overlap between both observing runs. Finally, the absorption technique
provides a high spatial resolution, equivalent to the angular size of the quasar continuum images and
corresponding to a few parsecs in the plane of FG0.89.

On the other hand, several limitations have to be kept in mind. First, the detection of molecular species
in our survey (for a given abundance) is obviously biased towards those that have {\em i)} a relatively
high dipole moment, {\em ii)} a relatively low partition function, and {\em iii)} low energy transitions
within the frequency coverage (57--94~GHz, rest-frame). The molecular inventory is then limited by a
combination of these properties.
Second, the angular resolution of our observations is insufficient to resolve the background continuum
emission. While the different absorption components are well-resolved kinematically, this introduces an
uncertainty in the continuum illumination, hence also in the line opacities and column densities. We argue
however that the associated uncertainties are only $\sim$10\% for the bulk of the lines
(see \S.\ref{back-illumin}). Moreover, the spectral resolution is only a factor of a few smaller than the
width of the line ($\sim$20~km\,s$^{-1}$). This reduces the absorption line depths, likely causing opacities
to be slightly underestimated. We expect that this affects weak lines the most. The limited spectral
resolution also prevents us from measuring the saturation level of the HCO$^+$/HCN $1-0$ SW lines, which
could help us to derive the NE/SW flux ratio. Finally, we assume a source covering factor of unity for
each absorption component, implying that the derived optical depths and column densities are, strictly
speaking, lower limits.

\subsection{Isotopic ratios} \label{section-isotopicratios}

Isotopic abundances are directly affected by nucleosynthesis processes in stellar interiors, and isotopic
ratios are thus good probes of the chemical enrichment history of the Universe. For this purpose, molecular
absorption offers interesting prospects, owing to the rotational transitions of different isotopologues
being easily resolved at mm wavelengths, such that low column densities of the rare isotopologues are still
detectable e.g. for most isotopes of C, N, O, and S elements, and that abundances can be inferred directly
from absorption optical depths (if the line of the most abundant isotopologue is not saturated).

The $^{12}$C/$^{13}$C, $^{14}$N/$^{15}$N, $^{16}$O/$^{18}$O, $^{18}$O/$^{17}$O, and $^{32}$S/$^{34}$S
isotopic ratios were derived in FG0.89SW by \cite{mul06}, based on observations of the different
isotopologues of HCO$^+$, HCN, HNC, CS, and H$_2$S at 3 mm with the Plateau de Bure interferometer.
These ratios, particularly $^{18}$O/$^{17}$O and $^{32}$S/$^{34}$S, reveal significant differences
when compared to sources in the Local Universe (see Table~7 from \citealt{mul06}). Interestingly,
comparable values are found using the same absorption technique in the $z$=0.68 galaxy located in
front of the quasar B~0218+357 (Muller et al. in prep.). These results are essentially consistent
with the expectation that low-mass stars ($<$1.5~M$_\odot$) had no (or only a short) time to play
a major role in the gas enrichment of such young (a few Gyr old) galaxies. 

Our 7~mm survey now allows us to check the previous results obtained at 3~mm in FG0.89SW, in particular
for lines with different optical depths. In addition, time variability is not a concern for this 7~mm
dataset. Unfortunately, the limited spectral resolution of the survey prevents us from measuring ratios
across the lines, which is a thorough test of possible saturation, and we simply derive the isotopic
ratios from the ratios of integrated opacities.

\begin{figure}[h] \begin{center}
\includegraphics[width=8cm]{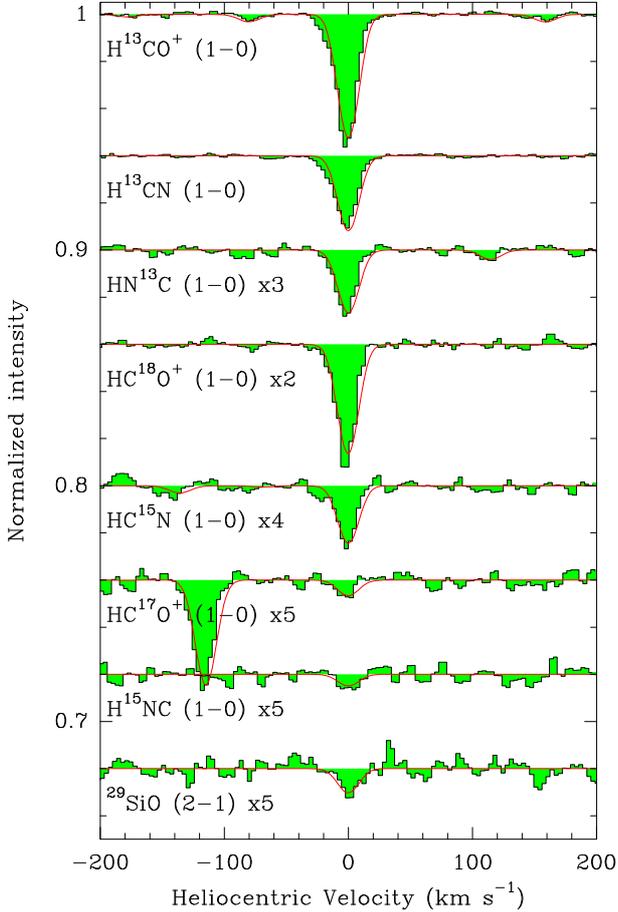}
\caption{Spectra of the different isotopologues of HCO$^+$, HCN, HNC, and SiO.}
\label{fig-isotopes}
\end{center} \end{figure}

The spectra of different isotopologues of HCO$^+$, HCN, HNC, and SiO as observed in our 7~mm survey are
shown in Fig.\ref{fig-isotopes}. Results for the corresponding isotopologue abundance ratios are given
in Table~\ref{tab-isotopologue-ratios}, where we also report the ratios previously measured at 3~mm.
Our final estimates for the various isotopic ratios in FG0.89SW are given in Table~\ref{tab-isotopicratios}.
The values obtained from 7~mm and 3~mm data are in very good agreement.
We hereafter discuss how we derived those ratios.

\subsubsection{Deuterium}

We do not detect lines from any of the deuterated species DCO$^+$/DCN, down to levels of a few 10$^{-4}$,
which is higher than the D/H cosmic abundance of 2.5$\times$10$^{-5}$ (e.g. \citealt{spe03}). This is
consistent with no strong fractionation enhancement, as expected in a relatively warm gas component
($T_{kin}$$\sim$50~K) with a moderate density, $n$(H$_2$) of the order of a few 10$^3$ cm$^{-2}$.

\subsubsection{Carbon}

It is not straightforward to go from the abundance ratio $^{12}$C$\star$/$^{13}$C$\star$, where C$\star$
is a given carbon-bearing molecule, to a value of the $^{12}$C/$^{13}$C isotopic ratio, because the measured
abundance ratio can be affected by opacity effects, chemical fractionation and/or selective photodissociation.
Hence, observations of several $^{12}$C- and $^{13}$C- bearing molecules are highly desirable.

The interstellar gas phase abundance of $^{13}$CO and $^{13}$C$^+$ relative to CO and C$^+$, can be affected by
the isotopic fractionation reaction
\begin{equation}
^{13}{\rm C}^+ + {\rm CO} \longleftrightarrow ^{13}{\rm CO} + {\rm C}^+ + 35~{\rm K}.
\label{eq-co-frac}
\end{equation}
\noindent In cold molecular clouds, $^{13}$C then becomes mostly trapped in $^{13}$CO, and the abundance of the
$^{13}$C$^+$ ion decreases. The $^{12}$C$\star$/$^{13}$C$\star$ ratio for species formed from CO should thus
reflect the $^{12}$C/$^{13}$C isotopic ratio, while for others $^{12}$C$\star$/$^{13}$C$\star$ $>$ $^{12}$C/$^{13}$C
should be observed. This was first investigated by \cite{lan84} for various carbon-bearing species. They found that
the $^{12}$C/$^{13}$C isotopic ratio should be within the limits set by the $^{12}$CO/$^{13}$CO and
H$_2^{12}$CO/H$_2^{13}$CO ratios. Since HCO$^+$ can be produced from both CO and other species, it can be enhanced
in either $^{12}$C or $^{13}$C, depending on the physical conditions.

The $^{12}$C$\star$/$^{13}$C$\star$ abundance ratios are listed in Table~\ref{tab-isotopologue-ratios} for
relevant species. The HCO$^+$/H$^{13}$CO$^+$ ratio might be underestimated because of the large opacity of
the $^{12}$C isotopologue. The opacity of HNC, on the other hand, is lower, and the HNC/H$^{13}$NC abundance
ratio might more accurately reflect the $^{12}$C/$^{13}$C elemental ratio. We do not detect transitions from
either the $^{13}$CCH or C$^{13}$CH isotopomers, which implies that the lower limits to the abundance ratios
are C$_2$H/$^{13}$CCH$>$58 and C$_2$H/C$^{13}$CH$>$43. \cite{sak10} found $^{13}$C abundance anomalies in the
$^{13}$C-isotopomer of C$_2$H toward TMC~1 and L1527, with the $^{13}$C-species underabundant relative to the
interstellar $^{12}$C/$^{13}$C ratio. In addition, the two carbon atoms do not appear to be equivalent in the
formation pathways of the molecule, as \cite{sak10} measured an abundance ratio C$^{13}$CH/$^{13}$CCH = 1.6
toward both sources. They conclude that carbon-chain molecules are not indicated to determine the
$^{12}$C/$^{13}$C elemental ratio, because of the positional differences and heavy $^{13}$C dilution. The
non-detection of H$_2^{13}$CO points to a ratio H$_2$CO/H$_2^{13}$CO$>$173, suggesting that fractionation
could affect our estimate of the $^{12}$C/$^{13}$C isotopic ratio. On the basis of all these results, we choose
to adopt the average ratio $^{12}$C/$^{13}$C=$35 \pm 11$, obtained from the HCO$^+$, HCN, and HNC species.

This value of $^{12}$C/$^{13}$C is less than half the ratio measured in the Solar System (89, \citealt{lod03}),
but closer to that derived in the local interstellar medium ($59 \pm 2$, \citealt{luc98}) and especially close
to that in the starburst galaxies NGC253 and NGC4945 (40--50, \citealt{hen93,cur01,wan04}). It is, however, much
lower than the value that would be expected in a poorly processed environment (e.g. \citealt{kob11}). We note
that \cite{mar10} used C$_2$H and CO isotopologues to revisit the $^{12}$C/$^{13}$C ratios in NGC253 and M82 and
found values larger than the ratios measured in previous studies. The interstellar $^{12}$C/$^{13}$C isotopic
ratio clearly remains difficult to determine.

Toward the NE component, we do detect neither the H$^{13}$CO$^+$ nor the H$^{13}$CN 1-0 absorptions, with lower
limits HCO$^+$/H$^{13}$CO$^+$$>$71 and HCN/H$^{13}$CN$>$41 (at 3~$\sigma$ level), respectively.
As the NE absorption is located at an even larger galactocentric distance than the SW, thus in a region likely
to be even less processed, we expect a $^{12}$C/$^{13}$C ratio larger than that toward FG0.89SW.

\subsubsection{Nitrogen} 

There are two ways of estimating the $^{14}$N/$^{15}$N isotopic ratios from our data: either from the
direct HCN/HC$^{15}$N and HNC/H$^{15}$NC abundance ratios, or, indirectly from the H$^{13}$CN/HC$^{15}$N
and HN$^{13}$C/H$^{15}$NC ratios, after correcting for the $^{12}$C/$^{13}$C isotopic ratio.

We obtain the ratios HCN/HC$^{15}$N=$192 \pm 11$, and HNC/H$^{15}$NC=340~$_{-70}^{+120}$. The former ratio might be
underestimated because of opacity effects on the HCN line, but the second ratio has a large uncertainty.
Using the $^{13}$C variants, we obtain H$^{13}$CN/HC$^{15}$N=$5.3 \pm 0.3$ and HN$^{13}$C/H$^{15}$NC=$7.3 _{-1.5}^{2.5}$.
The use of a large $^{12}$C/$^{13}$C ratio ($>$50) gives a $^{14}$N/$^{15}$N ratio 
close to the value observed in the local interstellar medium (=$237 \ _{-21}^{+27}$, \citealt{luc98}) and in the
Solar System (=272, \citealt{lod03}).
Using our estimate $^{12}$C/$^{13}$C=$35 \pm 11$ and the ratio H$^{13}$CN/HC$^{15}$N=$5.3 \pm 0.3$, we derive
$^{14}$N/$^{15}$N=$190 \pm 60$.

Whether this value actually reflects the true $^{14}$N/$^{15}$N isotopic ratio
is however uncertain if fractionation is an issue for nitrogen-bearing species (\citealt{rod08a, rod08b}).
The recent measurements of \cite{lis10} toward the two cold dense molecular clouds Barnard~1 and NGC1333 do not
suggest a high $^{15}$N enhancement.

We note that the $^{15}$N isotopologues of N$_2$H$^+$ are not detected, implying formally that $^{14}$N/$^{15}$N $> 90$.

\subsubsection{Oxygen}

As for $^{14}$N/$^{15}$N, the $^{16}$O/$^{18}$O isotopic ratio can be estimated either directly from the
HCO$^+$/HC$^{18}$O$^+$ abundance ratio, or indirectly from H$^{13}$CO$^+$/HC$^{18}$O$^+$. The former method
yields a ratio $^{16}$O/$^{18}$O=$50 \pm 1$, while the latter implies that $^{16}$O/$^{18}$=$80 \pm 25$.
Even considering that our value of $^{12}$C/$^{13}$C could be underestimated, the $^{16}$O/$^{18}$O
isotopic ratio derived in FG0.89SW is much lower than that in both the local interstellar medium (=$672
\pm 110$, \citealt{luc98}) and the Solar System (=499, \citealt{lod03}), in likely connection
with the youth of the galaxy.

Remarkably, the $^{18}$O/$^{17}$O isotopic ratio, measured directly from the H$^{18}$CO$^+$ and HC$^{17}$O$^+$
$J$=1-0 optically thin lines, also differs significantly from values for sources in the
Local Universe ($< 6$, see e.g. Table~7 of \citealt{mul06}).
While a value of $^{18}$O/$^{17}$O=$12 \ _{-3}^{+2}$ was measured from the $J$=2-1 lines, the marginal detection of
HC$^{17}$O$^+$ at 7~mm (Fig.\ref{fig-isotopes}) yields H$^{18}$CO$^+$/HC$^{17}$O$^+$=$20 ~_{-4}^{+7}$.
This new estimate of the $^{18}$O/$^{17}$O isotopic ratio confirms the large value previously
obtained from the $J$=2-1 lines at 3~mm.

\subsubsection{Silicon}

The detection of the $^{28}$SiO and $^{29}$SiO $J$=2-1 line allows us to estimate the isotopic ratio
$^{28}$Si/$^{29}$Si=11~$_{-2}^{+4}$.
Both SiO lines are optically thin, unless the source covering factor is much lower than unity.
This $^{28}$Si/$^{29}$Si ratio at $z$=0.89 is nearly half the value measured 
in the Solar System, 19.6 (\citealt{lod03}).

The situation is similar to that for the $^{32}$S/$^{34}$S isotopic ratio, estimated to
be $10.5 \pm 0.6$ from the CS/C$^{34}$S and H$_2$S/H$_2^{34}$S abundance ratios in FG0.89SW (\citealt{mul06}),
whereas it is 22 in the Solar System. 
Interestingly, the magnesium isotopic ratios were derived from UV-spectroscopy for a $z$=0.45 absorption line system
(\citealt{aga11}), and also suggest that there is a significant relative overabundance of heavy Mg isotopes.
Silicon and sulfur (and magnesium) are both produced by massive stars.
That the neutron-enriched isotopes appear to be more abundant at $z$=0.89 than in the Local Universe
might provide some constraint on their nucleosynthesis.

\begin{table}[ht]
\caption{Isotopologues abundance ratios toward FG0.89SW.} \label{tab-isotopologue-ratios}
\begin{center} \begin{tabular}{lcc}
\hline
\multicolumn{1}{c}{Abundance ratios} & @7 mm      & @3 mm    \\

\hline
DCO$^+$/HCO$^+$ & $< 7.1 \times 10^{-4}$ & -- \\
DCN/HCN         & $< 8.4 \times 10^{-4}$ & -- \\
DNC/HNC         &  --                 & $<0.015$ \\
DCO$^+$/H$^{13}$CO$^+$ & $< 0.015$  & -- \\
DCN/H$^{13}$CN         & $< 0.030$  & -- \\

\hline

HCO$^+$/H$^{13}$CO$^+$ & $21.2 \pm 0.2$ & $28 \pm 3$ \\
HCN/H$^{13}$CN         & $36.3 \pm 0.5$   & $40 \ _{-5}^{+7}$\\
HNC/HN$^{13}$C         & $47 \pm 2$   & $27 \pm 3$\\
H$_2$CO/H$_2^{13}$CO   & $> 173 ~(3\sigma)$ & -- \\

C$_2$H/$^{13}$CCH & $> 58 ~(3\sigma)$ & -- \\
C$_2$H/C$^{13}$CH & $> 43 ~ (3\sigma)$ & -- \\

\hline

HCN/HC$^{15}$N        & $192 \pm 11$ & -- \\
HNC/H$^{15}$NC        & $340 \ _{-70}^{+120}$ & $166 \ _{-58}^{+194}$ \\
H$^{13}$CN/HC$^{15}$N & $5.3 \pm 0.3$ & $4.2 \pm 0.3$ \\
HN$^{13}$C/H$^{15}$NC & $7.3 \ _{-1.5}^{+2.5}$ & $6 \ _{-2}^{+4}$\\

\hline
HCO$^+$/HC$^{18}$O$^+$        & $48 \pm 1$ & $53 \ _{-10}^{+16}$ \\
H$^{13}$CO$^+$/HC$^{18}$O$^+$ & $2.29 \pm 0.05$ & $2.01 \pm 0.07$ \\
HC$^{18}$O$^+$/HC$^{17}$O$^+$ & $19.6 \ _{-4}^{+7}$ & $12 \ _{-2}^{+3}$\\

\hline
$^{28}$SiO/$^{29}$SiO & $11 \ _{-2}^{+4}$ & -- \\

\hline
\end{tabular} \end{center} \end{table}

\begin{table}[ht] 
\caption{Isotopic ratios toward FG0.89SW.} \label{tab-isotopicratios}
\begin{center} \begin{tabular}{cccc}
\hline
Isotopic ratios    & 7 mm           & 3 mm    & Averaged \\
\hline
D/H                &  $< 7 \times10^{-4}$  & $<0.015$      & --\\
$^{12}$C/$^{13}$C   & $35 \pm 11$           & $32 \pm 6$     & $33 \pm 5$ \\
$^{14}$N/$^{15}$N   & $190 \pm 60$          & $143 \pm 30$   & $152 \pm 27$ \\
$^{16}$O/$^{18}$O   & $80 \pm 25$           & $66 \pm 12$     & $69 \pm 11$ \\
$^{18}$O/$^{17}$O   & $20 \ _{-4}^{+7}$       & $12 \ _{-2}^{+3}$ &  $13 \ _{-2}^{+3}$\\
$^{32}$S/$^{34}$S   & --                    & $10.5 \pm 0.6$   & $10.5 \pm 0.6$\\
$^{28}$Si/$^{29}$Si & $11 _{-2}^{+4}$         & --              & $11 _{-2}^{+4}$ \\

\hline
\end{tabular} \tablefoot{The values at 3 mm were rederived from \cite{mul06} data using the same methodology
as described in the text.} 
\end{center} \end{table}

\subsection{Comparative chemistry} \label{section-comparochemistry}

It is interesting to compare the molecular abundances in the $z$=0.89 galaxy with those measured
for various sources in the Local Universe, to characterize the type of clouds and chemistry. For
this purpose, we selected a sample of archetype sources with a large number of detected molecular
species, usually covered by large spectral surveys. Molecular abundances in circumstellar envelopes
around evolved stars, such as IRC+10216, were not included because of the distinctive chemical
segregation within the envelope. We preferred to use data obtained from a single team/telescope,
for the sake of homogeneity and to minimize beam effects.

Despite the low density and poor shielding from the interstellar radiation field, a variety of 
molecules is already present in Galactic diffuse clouds ($A_{V} < 1$): about 15 different species have
been detected so far (see e.g. \citealt{lis08}), and the limit of their chemical complexity
remains unclear. We mention here that column densities in diffuse clouds are most often
determined through mm-wave absorption toward extragalactic continuum sources, similar to
the work presented in this paper.

Translucent clouds (2$<$$A_{V}$$<$5) are expected to have a richer chemistry, with about 30 molecules
detected so far, including highly unsaturated and very reactive hydrocarbon chains. We adopt the
molecular abundances compiled by \cite{tur00a}.

Cold dark clouds are the sites of low-mass star formation, where an intricate gas-phase ion-molecule
interstellar chemistry takes place. We adopt the abundances determined by \cite{ohi92} toward TMC-1,
which is a prototype of dark clouds.

Hot and dense cores, associated with massive-star formation harbor the most complex interstellar
chemistry. There, the chemistry is characterized by the evaporation of dust grains, releasing large
molecules in the gas phase. We adopt the abundances derived by \cite{num00} in the region Sgr~B2N
toward the Galactic center as reference.

Finally, we also include in our comparison three extragalactic sources for which molecular
abundances have been derived for a significantly large number of species: the nuclear regions
of the starburst galaxies NGC253 and NGC4945 (\citealt{mar06} and \citealt{wan04}, respectively),
as well as the star forming region N113 in the Large Magellanic Cloud (LMC, \citealt{wan09}).

\begin{figure*}[h] \begin{center}
\includegraphics[width=\textwidth]{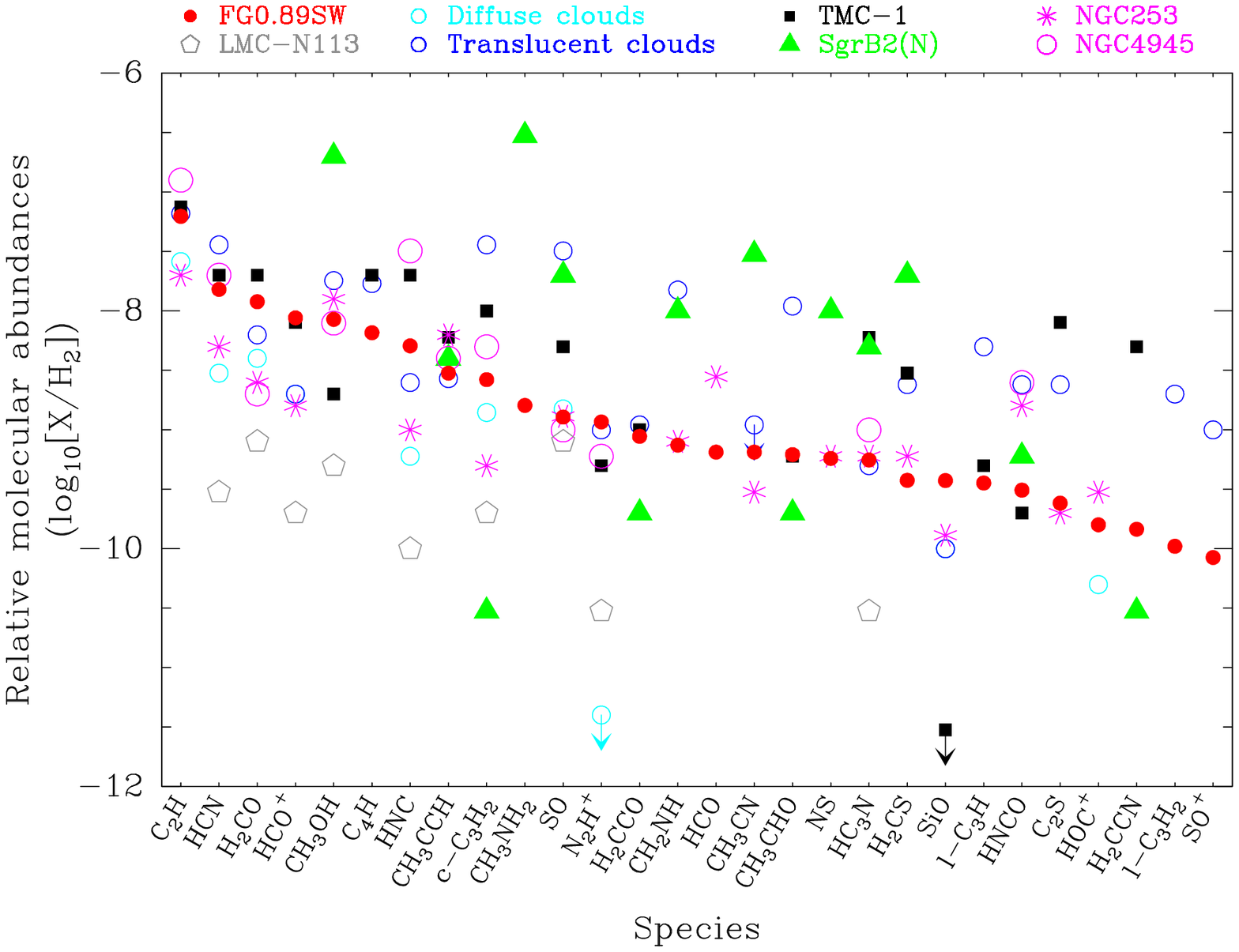}
\caption{Comparison of relative molecular abundances in FG0.89SW (from most to less abundant),
diffuse clouds (\citealt{luc00a,luc00b,lis01,luc02,lis04}), translucent clouds (\citealt{tur00a}),
TMC-1 (\citealt{ohi92}), Sgr~B2(N) (\citealt{num00}), LMC-N113 (\citealt{wan09}), and
the nuclear region of the starburst galaxies NGC253 (\citealt{mar06}) and NGC4945 (\citealt{wan04}).}
\label{fig-abundances}
\end{center} \end{figure*}

\begin{figure*}[h] \begin{center}
\includegraphics[width=\textwidth]{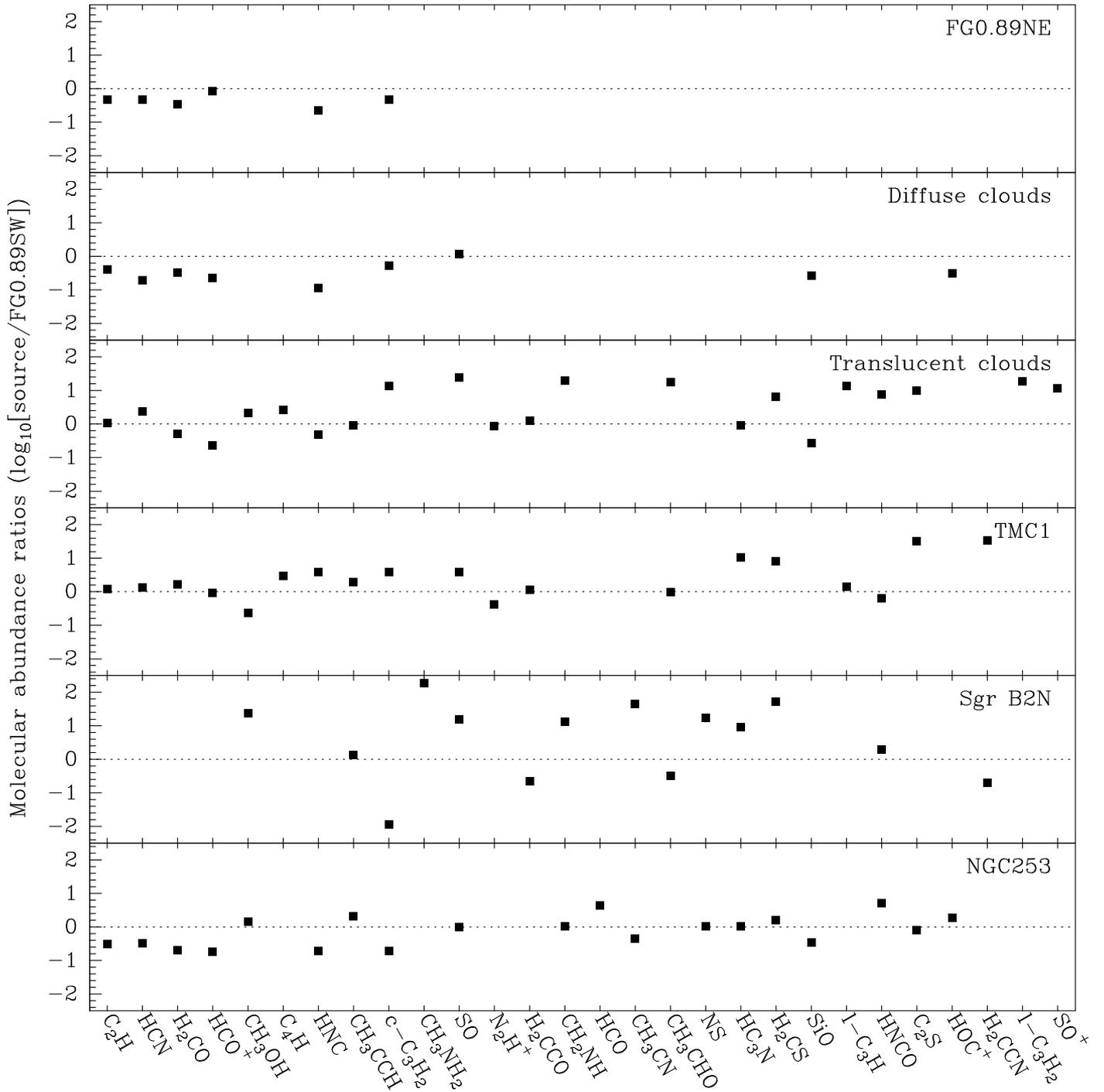}
\caption{Ratios of molecular abundances of FG0.89SW to other sources, detailed by species.}
\label{fig-ratio-XH2}
\end{center} \end{figure*}

The comparison of abundances derived in FG0.89SW with those observed toward these various sources
is illustrated in Fig.\ref{fig-abundances}. Molecular abundance ratios, normalized to FG0.89SW and
detailed by species, are also given in Fig.\ref{fig-ratio-XH2}. Several species in our survey, such
as HCO$^+$, HCN, C$_2$H, and c-C$_3$H$_2$ are commonly observed in Galactic diffuse clouds. More complex
molecules (e.g. carbon chains) are only seen toward cold dark clouds or in warm and dense clouds.

Comparing molecular abundances relies on the validity of various hypotheses. First, molecules have to
be co-spatial and share the same source covering factor. The size of the SW and NE compact continuum
components corresponds to a few pc in the plane of the $z$=0.89 galaxy. We assume a common source 
covering factor of unity for all molecules. However, the absorbing gas could be the mix of (most
likely several) dense cores embedded in a more diffuse component within the continuum beam
illumination. Our derived column densities and molecular abundances are thus lower limits. In addition,
opacity effects, if not correctly taken into account, could easily lead to erroneous molecular
abundances. For this reason, we do not use the abundances of HCO$^+$, HCN, and C$_2$H from \cite{num00}
toward Sgr~B2N, as they are probably largely underestimated. Finally, a reference species needs to be
assigned for normalization, in order to compare the relative molecular abundances. We choose to
normalize the column densities to that of molecular hydrogen, as commonly done in the literature.
For this, we assume a value $N$(H$_2$)=$2\times 10^{22}$~cm$^{-2}$ toward the SW component, as
estimated by \cite{wik98}. 
This value is consistent with the measurement of $N$(H)=$(3.5 \pm 0.6)$$\times$10$^{22}$~cm$^{-2}$
obtained from X-ray absorption by \cite{mat97} assuming that all absorption is due to the SW component.
The absorbing gas is indeed mostly molecular along this line of sight (\citealt{che99}), so that
$N$(H$_2$) should be close to 2$\times$$N$(H). The NE line of sight, on the other hand, intercepts
the disk of the $z$=0.89 galaxy at a galactocentric distance of $\sim$4~kpc. There, the extinction
$A_{V}$ is roughly one order of magnitude lower than toward the SW image of the quasar (e.g.
\citealt{fal99}, see also Fig.\ref{fig-hst}), and the column density of molecular gas is likely to
be similarly lower. We assume a H$_2$ column density of 1$\times$10$^{21}$ cm$^{-2}$ toward FG0.89NE.
This choice makes the fractional abundances similar to those in Galactic diffuse clouds 
(Table~\ref{tab-abundances} and Fig.\ref{fig-ratio-XH2}).
While the normalization to H$_2$ introduces an uncertainty in the absolute molecular abundances,
abundance ratios for a given source are not affected, and should therefore bear greater significance
(again provided that the species are co-spatial and have an identical source covering factor).

\begin{figure}[h] \begin{center}
\includegraphics[width=8cm]{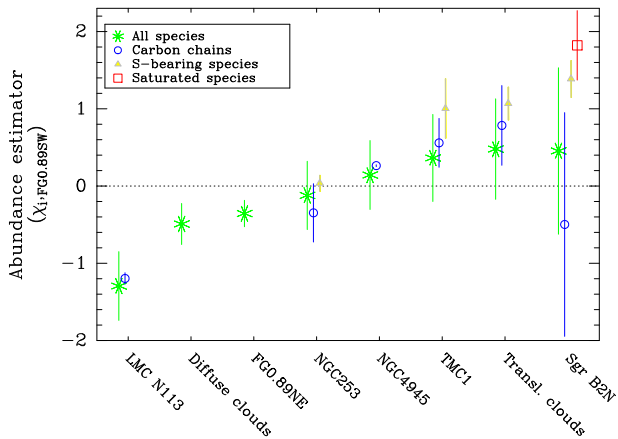}
\caption{Comparison of the abundance estimators $\chi_{ij}$ as defined in Eq.\ref{eq-chiij},
for different sources and sub-samples of molecular species (see text), with respect to FG0.89SW.}
\label{fig-comparochemistry}
\end{center} \end{figure}

To compare the overall trend between the various sources, and inspired by \cite{mar06}, we have
calculated an abundance estimator $\chi_{ij}$, defined as the average of the logarithm of abundance
ratios between two sources $i$ and $j$
\begin{equation} \label{eq-chiij}
\chi_{ij} = \frac{1}{N_{mol}} \sum_{k=1}^{N_{mol}} {\rm log_{10}} \left ( \frac{X_{i,k}}{X_{j,k}} \right ),
\end{equation}
where $X_{i,k}$ stands for the abundance relative to H$_2$ of molecule $k$ in source $i$, and the
sum runs over all $N_{mol}$ molecules detected in common for the two sources $i$ and $j$.
A dispersion in the estimator was also calculated to reflect potential large variations in the abundance
ratios between species. We do not take into account non-detections in this estimator.
Abundance estimators were also calculated in the same manner for the following sub-samples of molecules:
carbon chains (i.e. with more than two chained carbon atoms, such as l-C$_3$H, C$_4$H, c-C$_3$H$_2$,
l-C$_3$H$_2$, and HC$_3$N), sulfur-bearing species (i.e. SO, NS, H$_2$CS, C$_2$S, and SO$^+$), and
saturated molecules (CH$_3$OH and CH$_3$NH$_2$), presumably all formed on dust grains.
The value of $\chi_{ij}$ was calculated between FG0.89SW and the other sources only if the number of
common species for two objects was $\ge$2. The abundance estimators are given in Table~\ref{tab-Xindex}
and illustrated in Fig.\ref{fig-comparochemistry}.

On the basis of the abundance estimators including all species, we obtain a basic sequence of increasing
molecular abundances such as LMC $<$ diffuse clouds $<$ FG0.89NE $<$ NGC253 $<$ FG0.89SW $<$ NGC4945 $<$
TMC~1 $<$ translucent clouds $<$ Sgr~B2N, indicating, unsurprisingly, that the chemical abundances
in FG0.89SW are in-between those in typical Galactic diffuse and translucent clouds.

The dispersions for Sgr~B2N are large, though it is clear that sulfur-bearing and especially saturated
species have much higher abundances in this molecular-rich object. In TMC~1 and translucent clouds, the
abundance of carbon chains is globally a factor of a few higher than in FG0.89SW, while that of
sulfur-bearing species is one order of magnitude higher.

The overall similarity between the molecular abundances of FG0.89SW and the nuclear region of the
starburst galaxy NGC253 is surprising. \cite{mar06} found good agreement between the molecular
abundances in NGC253 and those at the position Sgr~B2(OH) in the Galactic center molecular cloud
Sgr~B2. Shocks are thought to play an important role in the heating and chemistry toward these
nuclear clouds, triggering the disruption of dust grains, and releasing the products of grain-surface
chemistry into the gas phase. It is tempting to invoke similar chemical processes in FG0.89SW, as
they could explain all of the similarities between the molecular abundances, the high kinetic
temperature of the gas, and the relatively large linewidth (see also discussions in \citealt{hen08,men08}).
Moreover, the elemental composition of the gas in both galaxies is expected to be dominated by the
nucleosynthesis products of massive stars, because of their shorter timescales. The major difference,
though, is that FG0.89SW clouds are located in a spiral arm at a galactocentric distance of about 2~kpc,
and gas densities are likely lower. Abundances in NGC4945 show some differences, albeit small, when
compared to those in NGC253. They are interpreted as different starburst evolutionary states
(\citealt{wan04, mar06}). The systematically lower molecular abundances in the LMC-N113 region are
attributed to a photon dominated region (PDR) in a nitrogen deficient environment (\citealt{wan09}).
Nevertheless, we emphasize that, because of the limited angular resolution of these extragalactic
observations, the corresponding molecular abundances reflect the average of the different types of
cloud gas properties and excitation conditions within the beam. This is much less the case for
absorption line studies toward quasars, where the spatial resolution is set by the small size of the
background continuum emission.

It will be interesting to investigate the chemical complexity toward FG0.89SW when searching for more
species, especially light hydrides and larger molecules. The molecular absorber located at $z$=0.68
in front of the quasar B~0218+357 would also be a suitable target for these investigations.


\begin{table*}[h] 
 \caption{Relative abundances [X]/[H$_2$] (10$^{-11}$).} \label{tab-abundances} 
 \begin{center} \begin{tabular}{lccccccccc} 
 \hline 
 \multicolumn{1}{c}{Species} & FG0.89SW & FG0.89NE & Diffuse & Translucent & TMC~1 & Sgr~B2(N) & LMC & NGC253 & NGC4945 \\ 
  &  &  & clouds  & clouds &  &  &  N113 &  &  \\ 
 \hline 
              C$_2$H& 6242.4&
 2924.5&
 2600.0&
 6600.0&
 7500.0&
--&
--&
 2000.0&
12600.0\\
                 HCN& 1520.4&
  733.0&
  300.0&
 3600.0&
 2000.0&
--&
   30.0&
  500.0&
 2000.0\\
             H$_2$CO& 1196.3&
  422.4&
  400.0&
  630.0&
 2000.0&
--&
   80.0&
  250.0&
  200.0\\
             HCO$^+$&  875.4&
  746.7&
  200.0&
  200.0&
  800.0&
--&
   20.0&
  160.0&
--\\
            CH$_3$OH&  849.9&
--&
--&
 1800.0&
  200.0&
20000.0&
   50.0&
 1260.0&
  790.0\\
              C$_4$H&  656.7&
--&
--&
 1700.0&
 2000.0&
--&
--&
--&
--\\
                 HNC&  509.2&
  116.8&
   60.0&
  250.0&
 2000.0&
--&
   10.0&
  100.0&
 3200.0\\
           CH$_3$CCH&  300.0&
--&
--&
  270.0&
  600.0&
  400.0&
--&
  630.0&
  400.0\\
        c-C$_3$H$_2$&  264.5&
  125.0&
  140.0&
 3600.0&
 1000.0&
    3.0&
   20.0&
   50.0&
  500.0\\
        CH$_3$NH$_2$&  160.6&
--&
--&
--&
--&
30000.0&
--&
--&
--\\
                  SO&  128.1&
--&
  150.0&
 3200.0&
  500.0&
 2000.0&
   80.0&
  130.0&
  100.0\\
          N$_2$H$^+$&  116.6&
  -33.4&
   -0.4&
  100.0&
   50.0&
--&
    3.0&
--&
   60.0\\
            H$_2$CCO&   88.3&
--&
--&
  110.0&
  100.0&
   20.0&
--&
--&
--\\
            CH$_2$NH&   74.5&
--&
--&
 1500.0&
--&
 1000.0&
--&
   80.0&
--\\
                 HCO&   65.0&
--&
--&
--&
--&
--&
--&
  280.0&
--\\
            CH$_3$CN&   65.0&
--&
--&
 -110.0&
--&
 3000.0&
--&
   30.0&
--\\
           CH$_3$CHO&   62.0&
--&
--&
 1100.0&
   60.0&
   20.0&
--&
--&
--\\
                  NS&   57.5&
--&
--&
--&
--&
 1000.0&
--&
   60.0&
--\\
             HC$_3$N&   55.9&
--&
--&
   50.0&
  600.0&
  500.0&
    3.0&
   60.0&
  100.0\\
             H$_2$CS&   37.6&
--&
--&
  240.0&
  300.0&
 2000.0&
--&
   60.0&
--\\
                 SiO&   37.5&
--&
   10.0&
   10.0&
   -0.3&
--&
--&
   13.0&
--\\
            l-C$_3$H&   35.7&
--&
--&
  500.0&
   50.0&
--&
--&
--&
--\\
                HNCO&   31.1&
--&
--&
  240.0&
   20.0&
   60.0&
--&
  160.0&
  250.0\\
              C$_2$S&   24.2&
--&
--&
  240.0&
  800.0&
--&
--&
   20.0&
--\\
             HOC$^+$&   15.9&
--&
    5.0&
--&
--&
--&
--&
   30.0&
--\\
            H$_2$CCN&   14.6&
--&
--&
--&
  500.0&
    3.0&
--&
--&
--\\
        l-C$_3$H$_2$&   10.5&
--&
--&
  200.0&
--&
--&
--&
--&
--\\
              SO$^+$&    8.4&
--&
--&
  100.0&
--&
--&
--&
--&
--\\
\hline \end{tabular} \end{center} 
\tablefoot{Negative values represent an upper limit. H$_2$ column densities of 2$\times$10$^{22}$ and 1$\times$10$^{21}$~cm$^{-2}$ were assumed for FG0.89SW and FG0.89NE, respectively.} 
\tablebib{Diffuse clouds: \cite{luc00a,luc00b,lis01,luc02,lis04}; Translucent clouds: \cite{tur00a}; 
TMC-1: \cite{ohi92}; SgrB2(N): \cite{num00}; LMC-N113: \cite{wan09}, using $N($H$_2$)=$2\times10^{22}$ cm$^{-2}$; NGC253: \cite{mar06}; NGC4945: \cite{wan04}.} 
\end{table*}

\begin{table*}[h] 
\caption{Abundance estimator.} \label{tab-Xindex} 
\begin{center} \begin{tabular}{lcccc} 
\hline 
 Source & All     & Carbon & S-bearing & Saturated \\ 
        & species & chains & species   & species  \\ 
\hline 
FG0.89SW & 0 & 0 & 0 & 0  \\ 
            LMC~N113&
  -1.3(   0.4)~[ 9]&
  -1.2(   0.1)~[ 2]&
--~[ 1]&
--~[ 1]\\
      Diffuse~clouds&
  -0.5(   0.3)~[ 9]&
--~[ 1]&
--~[ 1]&
--~[ 0]\\
            FG0.89NE&
  -0.4(   0.2)~[ 6]&
--~[ 1]&
--~[ 0]&
--~[ 0]\\
              NGC253&
  -0.1(   0.4)~[19]&
  -0.3(   0.4)~[ 2]&
   0.0(   0.1)~[ 4]&
--~[ 1]\\
             NGC4945&
   0.1(   0.4)~[11]&
   0.3(   0.0)~[ 2]&
--~[ 1]&
--~[ 1]\\
      Transl.~clouds&
   0.5(   0.7)~[22]&
   0.8(   0.5)~[ 5]&
   1.1(   0.2)~[ 4]&
--~[ 1]\\
                TMC1&
   0.4(   0.6)~[19]&
   0.6(   0.3)~[ 4]&
   1.0(   0.4)~[ 3]&
--~[ 1]\\
             Sgr~B2N&
   0.5(   1.1)~[13]&
  -0.5(   1.4)~[ 2]&
   1.4(   0.2)~[ 3]&
   1.8(   0.4)~[ 2]\\
\hline \end{tabular} \tablefoot{ 
The abundance estimators were calculated according to Eq.\ref{eq-chiij}. The dispersion is given in parenthesis and the number of molecules detected in common between FG0.89SW and other sources for each sample is given within brackets. 
} \end{center} \end{table*}

\subsection{Kinematics} \label{section-velo}

\subsubsection{Velocity centroids and line profiles}

\begin{figure}[h] \begin{center}
\includegraphics[width=8cm]{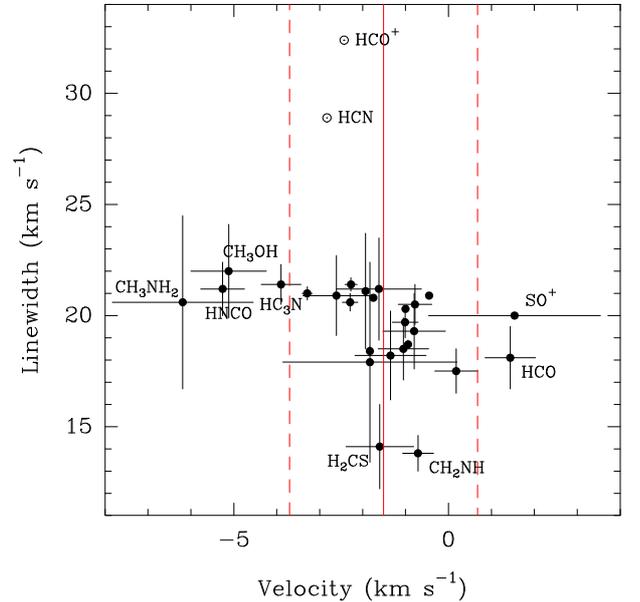}
\caption{Diagram of the FWHM versus velocity centroids, derived from Gaussian fits to the lines toward
FG0.89SW. The average velocity ({\em full line}) and 3$\sigma$ dispersion ({\em dashed lines}), calculated
from 22 species (see text), are indicated. The two points with a linewidth larger than 25~km\,s$^{-1}$
correspond to HCO$^+$ and HCN, the lines of which are likely saturated.}
\label{fig-velo-dv}
\end{center} \end{figure}

The limited velocity resolution of our survey (6--10 km\,s$^{-1}$, although slightly improved after the
Doppler correction) prevents us from conducting a detailed analysis of the kinematics and line profiles.
Nevertheless, the high S/N of most of the lines allows us to determine the velocity centroids of a large
number of species from Gaussian fitting, with an uncertainty of the order of one km\,s$^{-1}$ or better.
Inter-species velocity offsets can then be tested statistically. The results are shown in
Fig.\ref{fig-velo-dv}, where we show the FWHM versus the velocity centroid derived from Gaussian fits for
species seen toward the SW absorption.

Excluding the strong and likely saturated absorption of HCO$^+$ and HCN, we obtain an average velocity of
$-1.78 \pm 1.83$~km\,s$^{-1}$ toward FG0.89SW. Few species, namely CH$_3$NH$_2$, CH$_3$OH, l-C$_3$H, HNCO,
SO$^+$, and HC$_3$N, have velocities that differ from this average by more than 1~$\sigma$. These kinematical
differences might be due, e.g., to chemical segregation along the line of sight, different source covering
factors per species, excitation conditions, asymmetric line profiles, or a limited velocity resolution and/or
S/N of the observations. Another interpretation is explored in the next section. After removing the outliers
listed above, we obtain a new estimate of the average velocity: $-1.51 \pm 0.73$~km\,s$^{-1}$, the dispersion
dropping by a factor of more than two with respect to the first value including outliers.

\subsubsection{Constraints on the variation in the fundamental constants}

We now discuss the interpretation of the velocity offsets in terms of variations in the fundamental
constants of physics, and more particularly, in the proton-to-electron mass ratio.

Searches for observational evidence of any secular variation in the fundamental constants (FC) of
physics has attracted much attention lately. Any discovery of such evidence would indeed contradict
the Standard Model in the invariance principle, and open new horizons in physics research. We note
that several theories predict that the fundamental constants vary, and we refer the interested reader
to a review of this field by \cite{uza11}. A description of the various methods that have been applied
to constrain the variation in the fundamental constants to date, their results, and potential
systematics are also given in this review.

Spectroscopic observations of absorption lines toward high-redshift quasars naturally offer very long time
lapse, possibly longer e.g. than the age of the Solar System, that could allow even small variation rates
to be measured. While different methods have been developed, they all consist of a change in the frequency
of atomic or molecular transitions induced by a potential variation of the (dimensionless) fine structure
constant ($\alpha$), proton-to-electron mass ratio ($\mu$=$m_{p}/m_{e}$), and nuclear g-factor ($g_{n}$).
One can write this change as
\begin{equation}
\frac{\Delta \nu}{\nu} = K_\alpha \frac{\Delta \alpha}{\alpha} + K_\mu \frac{\Delta \mu}{\mu} +
K_{g} \frac{\Delta g_{n}}{g_{n}} = \sum _{C=\alpha,\mu,g_n} K_C \frac{\Delta C}{C},
\end{equation}
\noindent where the $K_i$ coefficients reflect the sensitivity of a change in the frequency of the
transition $\nu$ to a change in the constant $C$=$\alpha$, $\mu$ and/or $g_n$. The comparison of
two different lines is then given by
\begin{equation} \label{eq-dv}
\frac{\Delta V}{c} = \frac{V_i - V_j}{c} = \sum _{C=\alpha,\mu,g_n} (K_{C,j}-K_{C,i})\frac{\Delta C}{C},
\end{equation}
\noindent where $\Delta V$ is the velocity difference between the lines $i$ and $j$ induced by changes in FCs.
From Eq.\ref{eq-dv}, it is clear that only the comparison of lines with different $K_C$ sensitivity coefficients 
yields a possibly non-null velocity shift. For example, one has $|K_\alpha|$$<<$1, $K_\mu$=$-$1 and $|K_g|$$<<$1
for rotational transitions. For hyperfine transitions, $K_\alpha$=$-$2, $K_\mu$=$-$1, and $K_g$=$-$1.
Therefore, the comparison of hyperfine and rotational lines is sensitive to a variation of $\Delta F/F$,
where $F$=$\alpha^2g_n$.


Inversion lines of ammonia have a high sensitivity to any variation in $\mu$, with $K_\mu$=$-$4.46, but
are nearly insensitive to changes in other constants (\citealt{fla07}). As a result, the comparison of
NH$_3$ inversion lines and other rotational transitions provide a test of $\frac{\Delta \mu}{\mu}$.
This method was used toward the quasars B~0218+357 and PKS~1830$-$211 (the only two known high-$z$ ammonia
absorbers) and yields no significant evidence of changes in $\mu$, with constraints on $\frac{\Delta \mu}{\mu}$
down to a few 10$^{-7}$ (\citealt{mur08,hen09,kan11}), corresponding to a velocity offset of the order of a
few tenths of km\,s$^{-1}$. Nevertheless, the method is limited by the following systematic effects: (1)
local velocity offsets between species; (2) non-LTE and saturation effects for NH$_3$ hyperfine strucure
(if hyperfine components are unresolved or poorly detected) and optically thick lines, respectively;
(3) time variations (if observations are taken at different epochs); and (4) changes in the continuum morphology
at significantly different frequencies.

\cite{hen09} measured a velocity of $-3.70 \pm 0.43$~km\,s$^{-1}$ for HC$_3$N, consistent with our value of
$-3.90 \pm 0.46$~km\,s$^{-1}$. They used this value as a reference to constrain the velocity shift of NH$_3$
inversion transitions, obtaining a 3~$\sigma$ upper limit on $\frac{\Delta \mu}{\mu}$ of 1.4$\times$$10^{-6}$.
Taking our average velocity (without outliers) as reference instead, we derive a new value
$\frac{\Delta \mu}{\mu}$=($-2.04 \pm 0.74$)$\times$$10^{-6}$ from the ammonia velocity shift, assuming no
time variations between the different observations and that the shift is indeed due to variations in $\mu$.
Taken at face value, the velocity dispersion of the different species in our survey introduces a significant
contribution as {\em kinematical noise}, potentially limiting constraints on $\frac{\Delta \mu}{\mu}$
to $\sim$2$\times$$10^{-6}$ using inversion lines of ammonia.

Interestingly, methanol was also found to have transitions with high sensitivity to change in the
proton-to-electron mass ratio (\citealt{jan11}). The crude offset between the velocity of methanol and the
average velocity is $\Delta V$=$-3.75 \pm 0.89$~km\,s$^{-1}$. Taking the sensitivity coefficient $K_{\mu}$=$-$7.4
for the methanol transition at 60.531477 GHz (\citealt{jan11}) and provided that the velocity offset is due only
to a variation in $\mu$, we obtain $\frac{\Delta \mu}{\mu}$=($-1.95 \pm 0.47$)$\times$$10^{-6}$, that is, a
statistical 4~$\sigma$ significant detection of a variation in $\mu$ at $z$=0.88582\footnote{That is, a look-back
time of 7.24~Gyr for $H_0$=71~km\,s$^{-1}$~Mpc$^{-1}$, $\Omega_M$=0.27, and $\Omega_\Lambda$=0.73.}, in agreement
with the ammonia result. However, since we detect only one methanol line in our survey, we consider this result
as only tentative. More conservatively, our observations give an upper limit
$|\frac{\Delta \mu}{\mu}|$$<$4$\times$$10^{-6}$, which, importantly, takes into account the velocity dispersion
of a large number of molecular species. More species and lines with different sensitivity coefficients should
be observed at high velocity resolution and high S/N to check the compatibility with variations in $\mu$. We note
that \cite{kan11} derived $\frac{\Delta \mu}{\mu}$=($-3.5 \pm 1.2$)$\times$$10^{-7}$ at $z$=0.68 from the
comparison of NH$_3$ inversion lines and rotational lines of two species, CS and H$_2$CO, seen in absorption
toward B~0218+357.

\section{Summary and conclusions} \label{Conclusion}

We have completed the first unbiased spectral survey toward an intermediate-redshift molecular absorber,
the $z$=0.89 galaxy located in front of the quasar PKS~1830$-$211. Observations, performed with the Australian
Telescope Compact Array, cover the 7~mm band from 30 to 50~GHz (equivalent to the rest-frame frequency interval
57--94 GHz), and reach a sensitivity of the order of a few times 10$^{-3}$ of the continuum level, at a
spectral resolution of 1~MHz.

The brightness of the background quasar, the large column density of absorbing molecular gas, the large
instantaneous instrumental bandwidth, the wide frequency coverage of our survey, and the frequency band most
sensitive to low energy transitions of numerous molecules, all concur in allowing us to achieve the first
large chemical inventory in the interstellar medium of a distant galaxy.

From these observations, we derive the following results:
\begin{itemize}

\item We detect a total of 28 molecular species toward the SW image of the quasar (of which more than half are
detected for the first time toward this target and eight are detected for the first time outside the Milky Way:
SO$^+$, l-C$_3$H, l-C$_3$H$_2$, H$_2$CCN, H$_2$CCO, C$_4$H, CH$_3$NH$_2$, and CH$_3$CHO), and six toward the NE image.

\item The rotation temperatures are close to the expected cosmic microwave background temperature, 5.14~K,
at $z$=0.89. This absorber, with strong absorption along two independent lines of sight, offers a unique
opportunity to establish a robust and accurate measurement of $T_{CMB}$ at $z$=0.89.

\item The fractional molecular abundances toward the SW absorption are found to be in-between those in typical
Galactic diffuse and translucent clouds.

\item Toward the NE absorption, where the column density of molecular gas is about one order of magnitude lower,
the fractional abundances are comparable with those in diffuse clouds.

\item The isotopic ratios of C, N, O, and Si at 7~mm confirm and complement the C, N, O, and S ratios previously
obtained by \cite{mul06} at 3~mm.
Significant differences are found from the solar and local ISM values,
which could be explained by the limited timescale for nucleosynthetic processing at $z$=0.89.
Deuterated species remain undetected down to a D/H ratio of $<$7$\times$10$^{-4}$.

\item We discover several new, but weak, velocity components, at $V$=$-60$, $-224$, and $-300$~km\,s$^{-1}$,
seen in the HCO$^+$ and HCN (1-0) lines.

\item The analysis of the velocity centroids yields a dispersion of nearly 2~km\,s$^{-1}$ toward FG0.89SW. Among the
species showing a significant offset from the average velocity, NH$_3$ (observed by \citealt{hen09}) and CH$_3$OH
both have a high sensitivity to variations in the proton-to-electron mass ratio $\mu$.
We derive a conservative upper limit of $|\frac{\Delta \mu}{\mu}|$$<$4$\times$$10^{-6}$ at $z$=0.89, which includes
the velocity dispersion (hence the uncertainty caused by potentially different spatial distributions)
of a large number of molecular species. Further observations of lines with different sensitivity coefficients
will be required to place stronger constraints on variations in $\mu$.

\end{itemize}

The new generation of instruments currently deployed (ALMA, EVLA, NOEMA) will offer new capabilities in terms of frequency
coverage, instantaneous bandwidth, sensitivity, and spectral resolution that will be of great benefit to
molecular absorption studies.
It will be interesting to investigate the chemistry and its complexity in this
distant $z$=0.89 galaxy, as well as toward other known molecular absorbers.

\begin{acknowledgement}

We wish to thank Phil Edwards for his help and to acknowledge additional Director's observing time
allocated in March 2010 to complete the project.
We are grateful to Jamie Stevens and Kate Randall for their help with observations.
We thank the referee, Rainer Mauersberger, and the editor, Malcolm Walmsley, for their comments on the manuscript,
and Claire Halliday for the language editing.
We thank Sergei Levshakov for pointing out a mistake in the upper limit to the $\mu$ variation in a previous
version of the manuscript.
This research has made intensive use of the Cologne Database for Molecular Spectroscopy and
Jet Propulsion Laboratory Molecular Database.
The Australia Telescope Compact Array is part of the Australia Telescope which is funded by the
Commonwealth of Australia for operation as a National Facility managed by CSIRO.
S.M. acknowledges the support of a NORDFORSK grant in 2009--2010.

\end{acknowledgement}

\Online

\begin{appendix}

\section{Spectral survey and line catalog}

The full spectrum, from 30 to 50~GHz, is available on-line as an ASCII file at the {\em Strasbourg Astronomical Data Center} (CDS).


\begin{figure*}
\includegraphics[width=\textwidth]{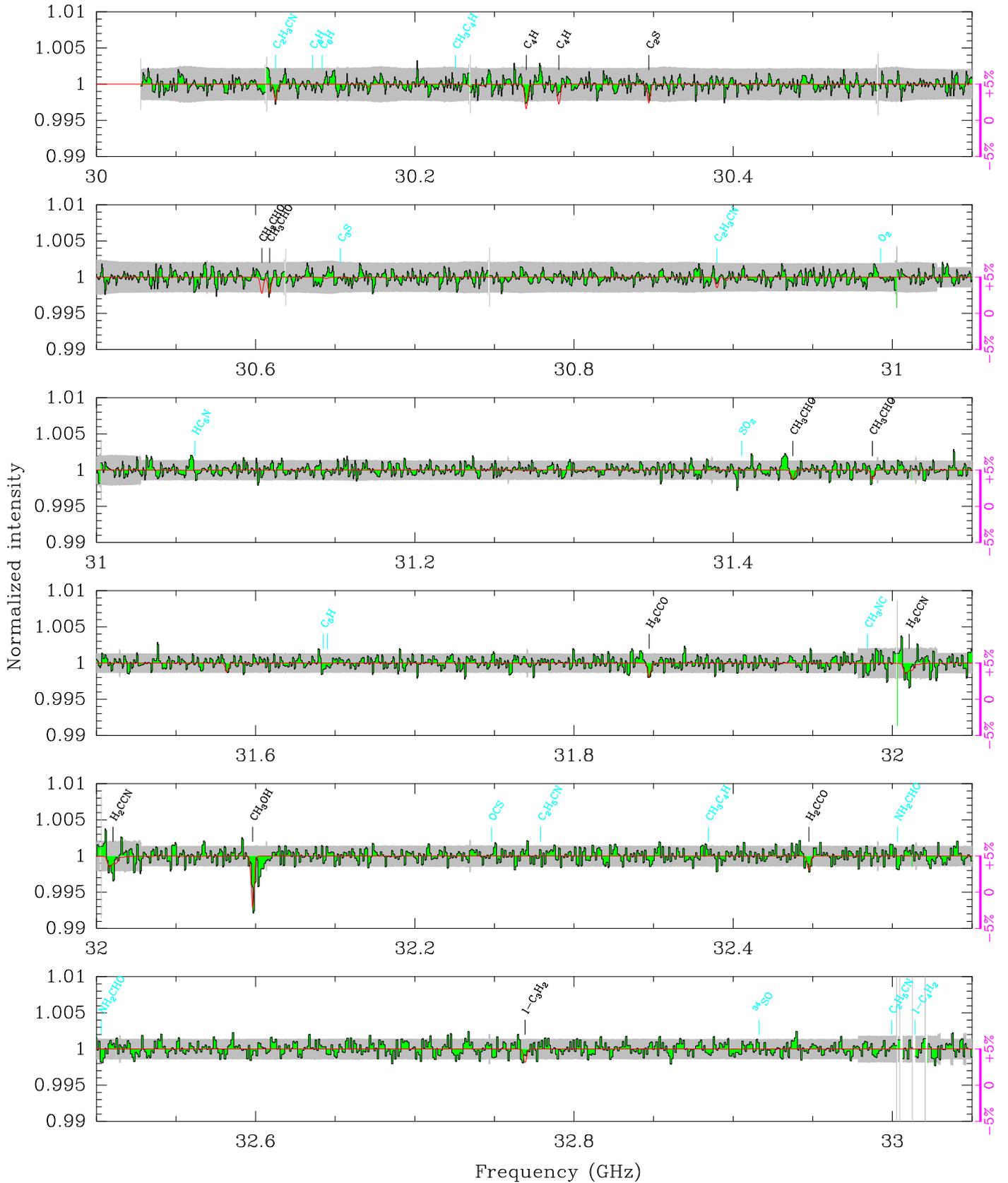}
\caption{Spectrum of the ATCA 7~mm survey toward PKS~1830$-$211. The grey envelope indicates the 3~$\sigma$ noise level.
The LTE synthetic model, assuming $T_{rot}$=5.14~K for all molecules, is overlaid in red. The spectrum in magenta,
offset at a level of 0.995, gives the relative differences between observations taken in September 2009 and March 2010,
normalized to the average spectrum. The corresponding scale is given on the right of each box ($\pm 5$\%).
Some non-detections are indicated in light blue.}
\label{fig-fullsurvey}
\end{figure*}

\begin{figure*} \addtocounter{figure}{-1}
\includegraphics[width=\textwidth]{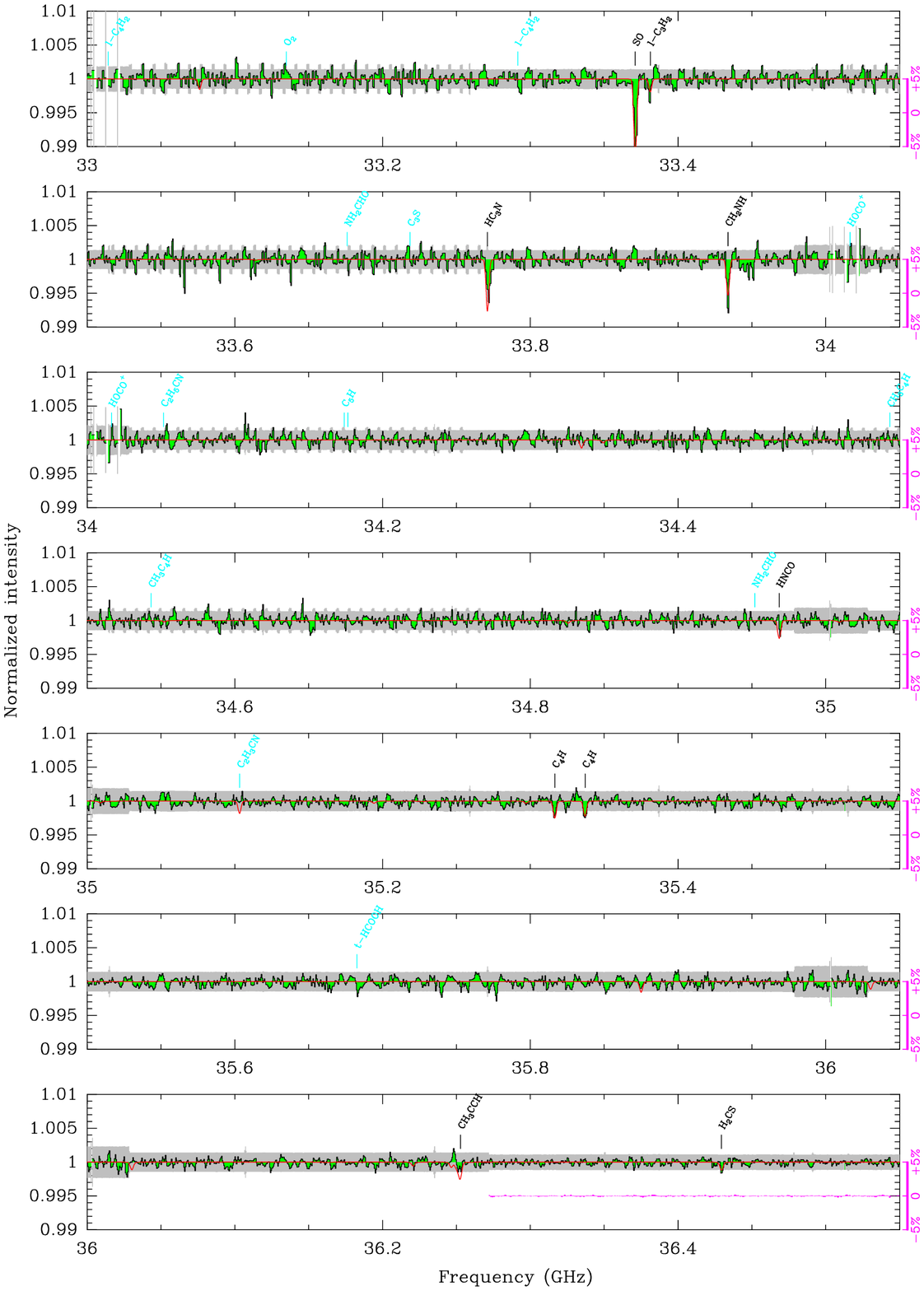}
\caption{{\em Continued.}}
\end{figure*}
\begin{figure*} \addtocounter{figure}{-1}
\includegraphics[width=\textwidth]{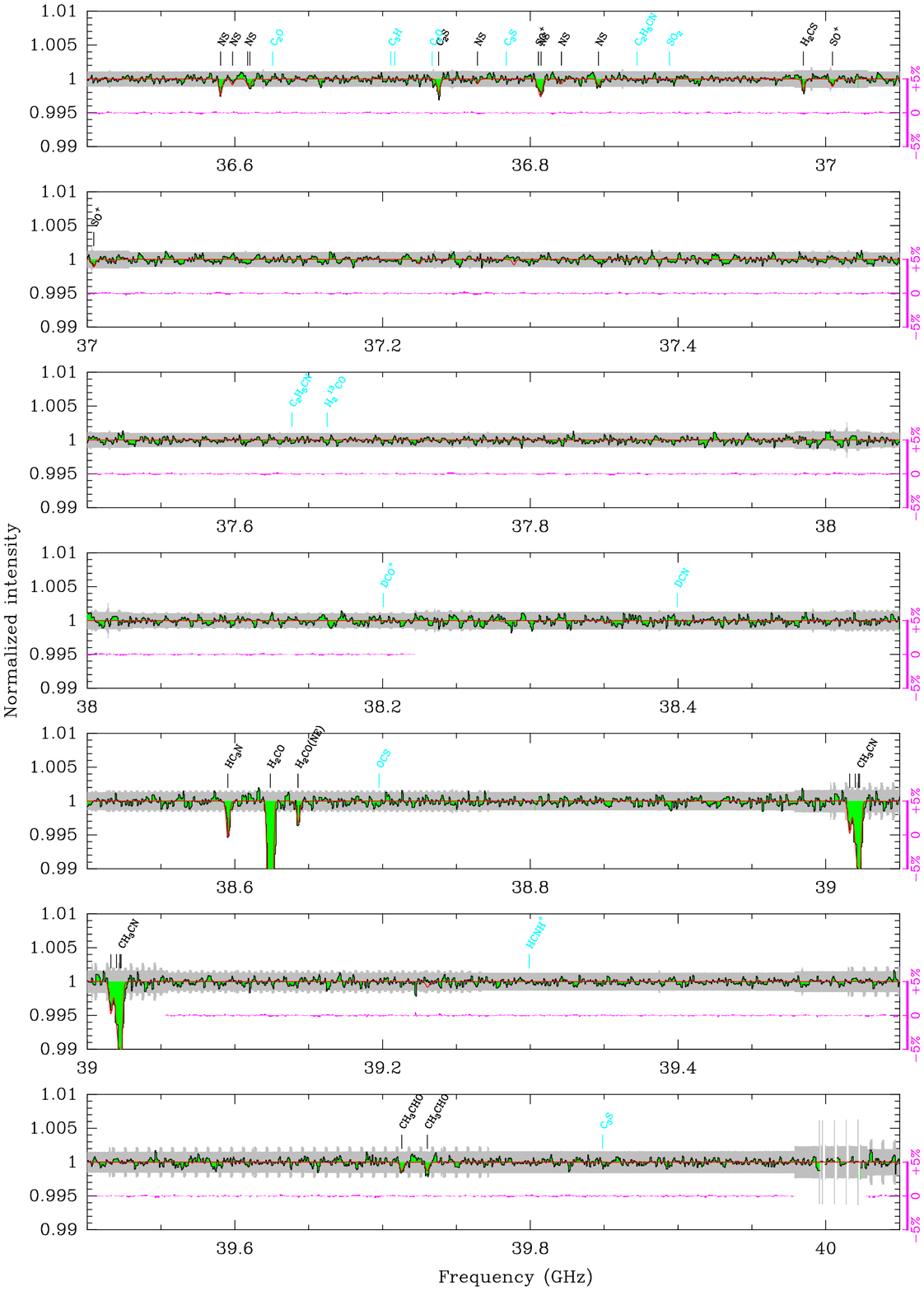}
\caption{{\em Continued.}}
\end{figure*}
\begin{figure*} \addtocounter{figure}{-1}
\includegraphics[width=\textwidth]{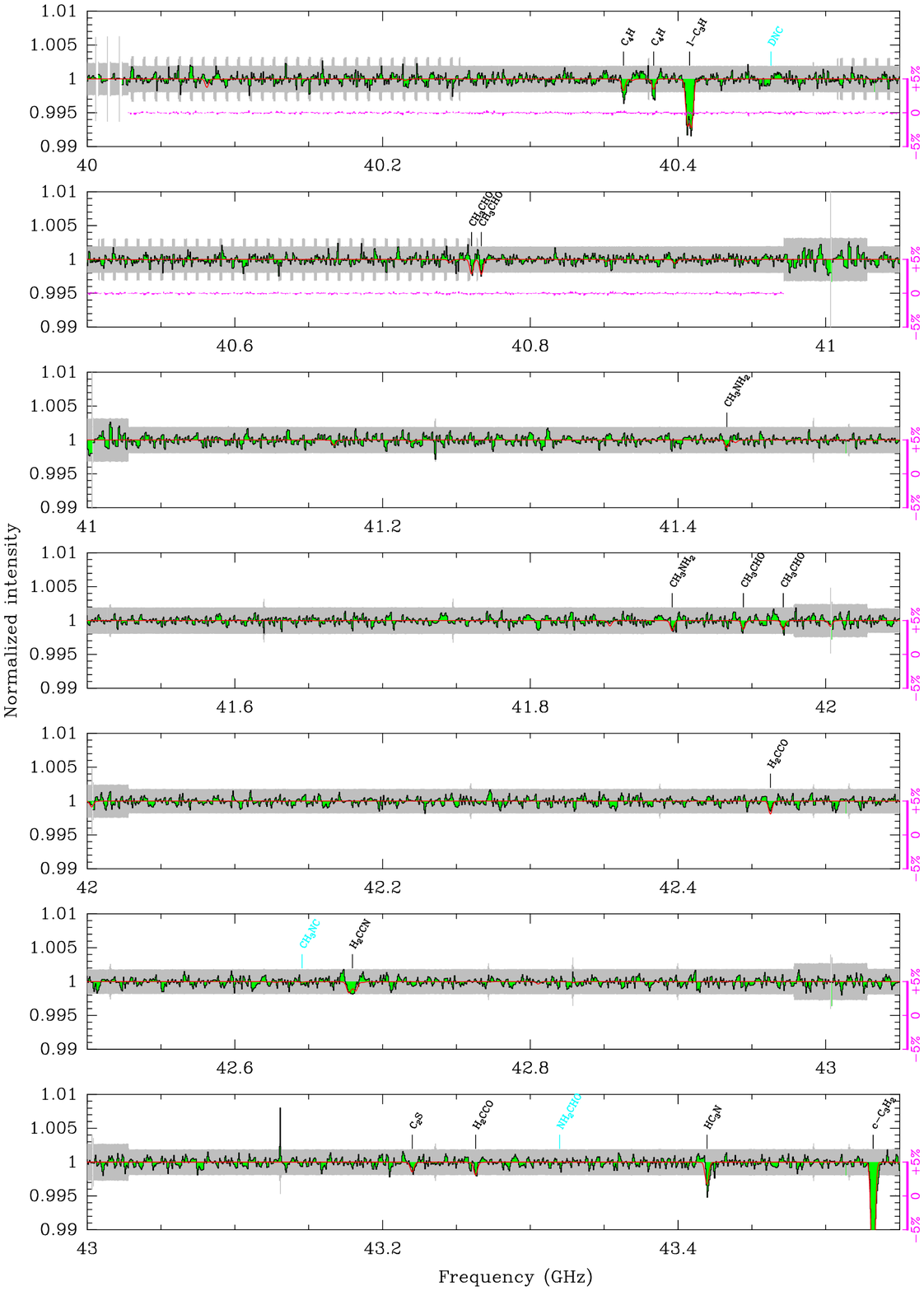}
\caption{{\em Continued.}}
\end{figure*}
\begin{figure*} \addtocounter{figure}{-1}
\includegraphics[width=\textwidth]{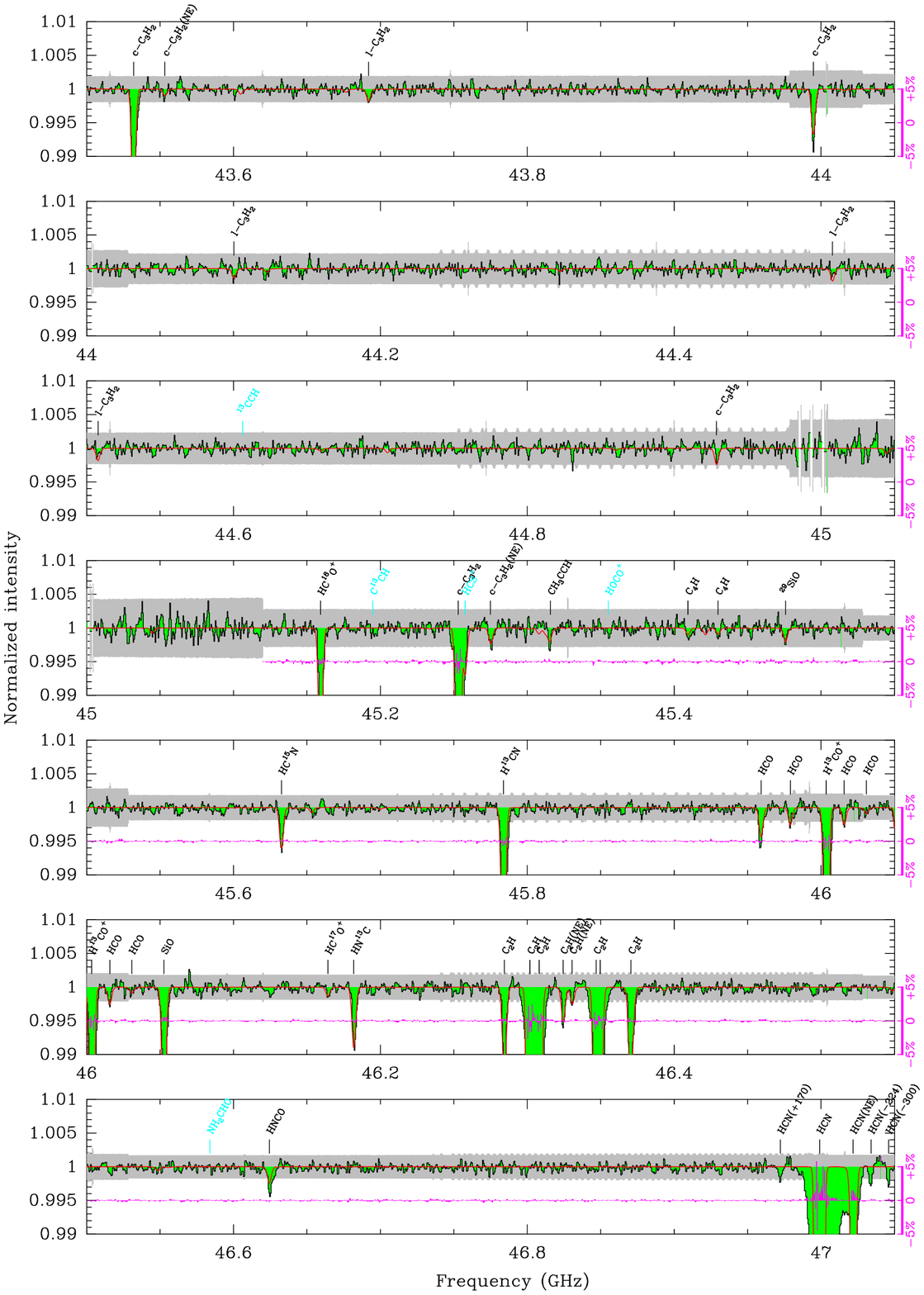}
\caption{{\em Continued.}}
\end{figure*}
\begin{figure*} \addtocounter{figure}{-1}
\includegraphics[width=\textwidth]{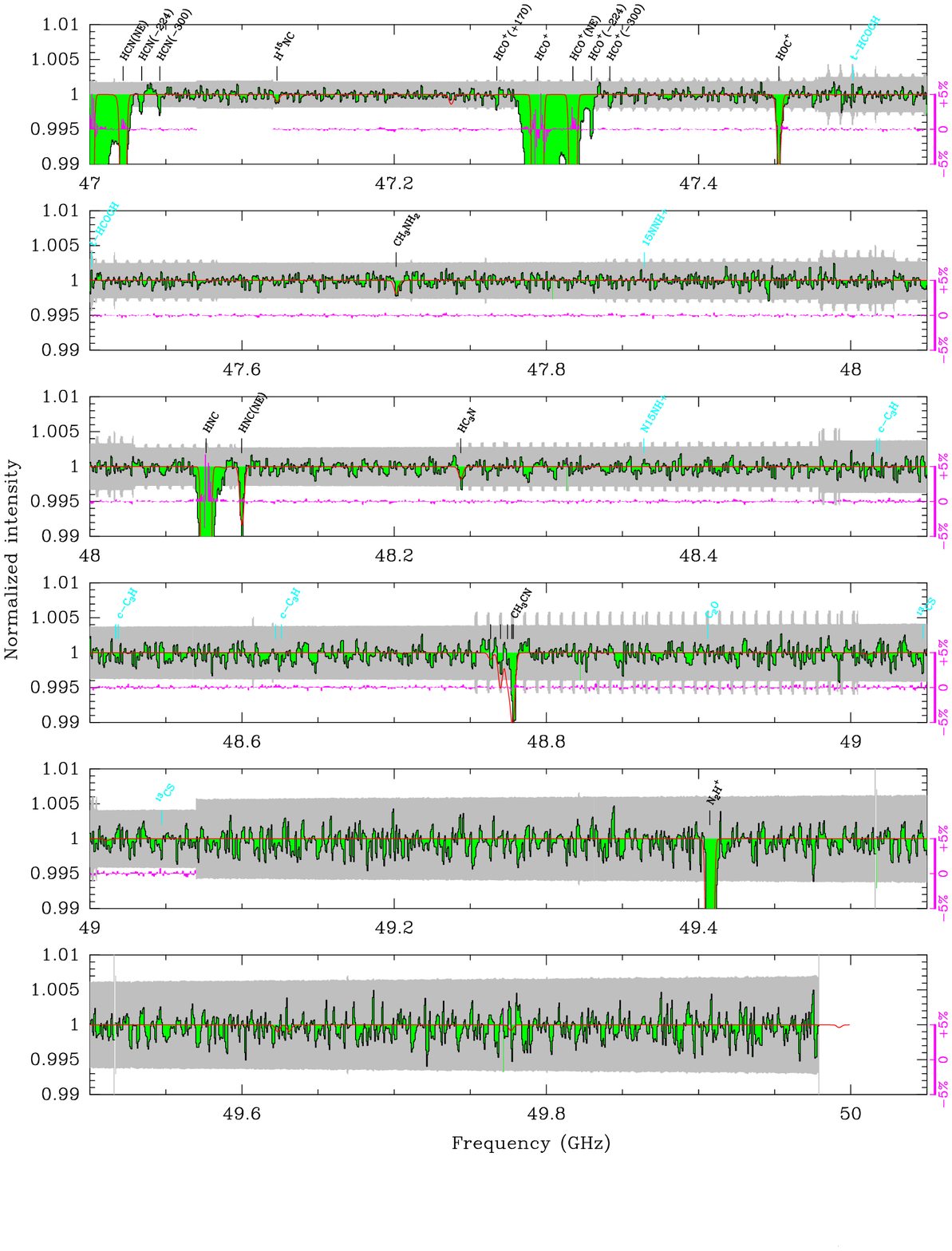}
\caption{{\em Continued.}}
\end{figure*}

\clearpage
\onecolumn
\begin{longtable}{llrrrrr}
\caption{Parameters and fitting results for lines detected toward FG0.89SW.} \label{tab-lines} \\
\hline
 
Rest Freq. & Transition  & \multicolumn{1}{c}{$S_{ul}$} & \multicolumn{1}{c}{$E_l/k_B$} & \multicolumn{1}{c}{V$_0$} & \multicolumn{1}{c}{$\Delta V$} & $\int \tau $d$V$\\
\multicolumn{1}{c}{(MHz)}      &             &          &  \multicolumn{1}{c}{(K)} & (km~s$^{-1}$) & (km~s$^{-1}$) & (10$^{-3}$ km~s$^{-1}$)\\
 
\hline
\endfirsthead
\caption{Continued.} \\
\hline
Rest Freq. & Transition  & \multicolumn{1}{c}{$S_{ul}$} & \multicolumn{1}{c}{$E_l/k_B$} & V$_0$ & $\Delta V$ & $\int \tau $d$V$\\
\multicolumn{1}{c}{(MHz)}      &             &          &  \multicolumn{1}{c}{(K)} & (km~s$^{-1}$) & (km~s$^{-1}$) & (10$^{-3}$  km~s$^{-1}$)\\
 
\hline
\endhead
 
\endfoot
87284.105&              C$_2$H~                             N=1-0 J=3/2-1/2 F=1-1& 0.04&  0.0&-0.45( 0.03)&20.90( 0.10)&  777.(    0.)\\
87316.898&              C$_2$H~                             N=1-0 J=3/2-1/2 F=2-1& 0.42&  0.0&--&--& 7625.(   21.)\\
87328.585&              C$_2$H~                             N=1-0 J=3/2-1/2 F=1-0& 0.21&  0.0&--&--& 3799.(    0.)\\
87401.989&              C$_2$H~                             N=1-0 J=1/2-1/2 F=1-1& 0.21&  0.0&--&--& 3805.(    0.)\\
87407.165&              C$_2$H~                             N=1-0 J=1/2-1/2 F=0-1& 0.08&  0.0&--&--& 1528.(    0.)\\
87446.470&              C$_2$H~                             N=1-0 J=1/2-1/2 F=1-0& 0.04&  0.0&--&--&  780.(    0.)\\
88631.602&                 HCN~                                             J=1-0& 1.00&  0.0&-2.83( 0.01)&28.90( 0.01)&68695.(  113.)\\
86339.922&          H$^{13}$CN~                                             J=1-0& 1.00&  0.0&-2.27( 0.14)&21.40( 0.30)& 1821.(   25.)\\
86054.966&          HC$^{15}$N~                                             J=1-0& 1.00&  0.0&--&--&  342.(   19.)\\
90663.568&                 HNC~                                             J=1-0& 1.00&  0.0&-1.75( 0.02)&20.80( 0.10)&24767.(   68.)\\
87090.825&          HN$^{13}$C~                                             J=1-0& 1.00&  0.0&--&--&  498.(   18.)\\
88865.715&          H$^{15}$NC~                                             J=1-0& 1.00&  0.0&--&--&   70.(   18.)\\
93173.392&          N$_2$H$^+$~                                             J=1-0& 1.00&  0.0&-3.29( 0.12)&21.00( 0.30)& 7341.(   84.)\\
89188.525&             HCO$^+$~                                             J=1-0& 1.00&  0.0&-2.43( 0.02)&32.40( 0.01)&67923.(  102.)\\
86754.288&      H$^{13}$CO$^+$~                                             J=1-0& 1.00&  0.0&-1.83( 0.07)&18.40( 0.20)& 3079.(   25.)\\
85162.223&      HC$^{18}$O$^+$~                                             J=1-0& 1.00&  0.0&--&--& 1310.(   28.)\\
87057.535&      HC$^{17}$O$^+$~                                             J=1-0& 1.00&  0.0&--&--&   69.(   17.)\\
89487.414&             HOC$^+$~                                             J=1-0& 1.00&  0.0&-0.78( 0.39)&20.50( 0.90)&  626.(   24.)\\
86670.760&                 HCO~           N$_K$=1$_{01}$-0$_{00}$ J=3/2-1/2 F=2-1& 0.42&  0.0& 1.44( 0.59)&18.10( 1.40)&  309.(   22.)\\
86708.360&                 HCO~           N$_K$=1$_{01}$-0$_{00}$ J=3/2-1/2 F=1-0& 0.25&  0.0&--&--&  115.(   19.)\\
86777.460&                 HCO~           N$_K$=1$_{01}$-0$_{00}$ J=1/2-1/2 F=1-1& 0.25&  0.0&--&--&  118.(   20.)\\
86805.780&                 HCO~           N$_K$=1$_{01}$-0$_{00}$ J=1/2-1/2 F=0-1& 0.08&  0.0&--&--&   47.(   19.)\\
63992.975&            CH$_2$NH~                                 1$_{01}$-0$_{00}$& 2.25&  0.0&-0.71( 0.36)&13.80( 0.80)&  301.(   16.)\\
72837.948&          H$_2$CO(p)~                     J$_{Ka,Kc}$=1$_{01}$-0$_{00}$& 1.00&  0.0&-1.00( 0.04)&20.30( 0.10)& 6061.(   22.)\\
78135.183&     CH$_3$NH$_2$(o)~                                 2$_{13}$-2$_{02}$&30.00&  6.7&-6.19( 1.64)&20.60( 3.90)&   60.(   21.)\\
79008.422&     CH$_3$NH$_2$(o)~                                 1$_{12}$-1$_{03}$&18.00&  2.4&--&--&   84.(   21.)\\
89956.068&     CH$_3$NH$_2$(p)~                                 1$_{10}$-1$_{01}$& 6.00&  2.1&--&--&  110.(   25.)\\
60531.489&            CH$_3$OH~                              1$_{01}$-2$_{-12}$ E& 1.80& 12.5&-5.26( 0.51)&21.20( 1.20)&  431.(   21.)\\
76198.726&            l-C$_3$H~                                         J=7/2-5/2& 3.85&  4.2& 0.18( 0.50)&17.50( 1.00)&  222.(    8.)\\
76199.928&            l-C$_3$H~                                         J=7/2-5/2& 2.85&  4.2&--&--&  167.(    0.)\\
76204.182&            l-C$_3$H~                                         J=7/2-5/2& 3.85&  4.2&--&--&  222.(    0.)\\
76205.103&            l-C$_3$H~                                         J=7/2-5/2& 2.85&  4.2&--&--&  167.(    0.)\\
82966.200&     c-C$_3$H$_2$(o)~                                 3$_{12}$-3$_{03}$& 2.93& 12.1&-0.94( 0.07)&18.70( 0.20)&  439.(   27.)\\
85338.893&     c-C$_3$H$_2$(o)~                                 2$_{12}$-1$_{01}$& 4.50&  2.3&--&--& 4013.(   34.)\\
59557.896&     c-C$_3$H$_2$(p)~                                 3$_{31}$-3$_{22}$& 1.37& 16.1&--&--&   71.(   15.)\\
82093.559&     c-C$_3$H$_2$(p)~                                 2$_{02}$-1$_{11}$& 1.37&  2.5&--&--& 1211.(   20.)\\
61797.009&     l-C$_3$H$_2$(o)~                                 3$_{13}$-2$_{12}$& 8.00& 16.3&-1.93( 0.00)&21.10( 2.60)&   72.(   17.)\\
62950.993&     l-C$_3$H$_2$(o)~                                 3$_{12}$-2$_{11}$& 8.00& 16.4&--&--&  132.(   18.)\\
82395.089&     l-C$_3$H$_2$(o)~                                 4$_{14}$-3$_{13}$&11.25& 19.3&--&--&  117.(   21.)\\
83165.345&     l-C$_3$H$_2$(p)~                                 3$_{03}$-2$_{02}$& 4.00&  6.0&--&--&   83.(   23.)\\
60366.000&         H$_2$CCN(o)~                                 3$_{03}$-2$_{02}$&54.00&  2.9&-1.83( 2.03)&17.90( 4.50)&  264.(   47.)\\
80486.000&         H$_2$CCN(o)~                                 4$_{04}$-3$_{03}$&72.00&  5.8&--&--&  191.(   30.)\\
85457.300&           CH$_3$CCH~                                 $J_K$=4$_0$-3$_0$&10.00&  8.2& 0.59( 1.54)&15.60( 4.00)&  130.(   44.)\\
73577.451&            CH$_3$CN~                                 J$_K$=4$_3$-3$_3$& 7.00& 69.6&-2.70( 0.00)&21.60( 0.00)&--\\
73584.543&            CH$_3$CN~                                 J$_K$=4$_2$-3$_2$& 6.00& 33.9&--&--&--\\
73588.799&            CH$_3$CN~                                 J$_K$=4$_1$-3$_1$& 7.50& 12.4&--&--&--\\
73590.218&            CH$_3$CN~                                 J$_K$=4$_0$-3$_0$& 8.00&  5.3&--&--&--\\
91958.726&            CH$_3$CN~                                 J$_K$=5$_4$-4$_4$& 3.60&123.1&--&--&--\\
91971.130&            CH$_3$CN~                                 J$_K$=5$_3$-4$_3$&12.80& 73.1&--&--&--\\
91979.994&            CH$_3$CN~                                 J$_K$=5$_2$-4$_2$& 8.40& 37.4&--&--&--\\
91985.314&            CH$_3$CN~                                 J$_K$=5$_1$-4$_1$& 9.60& 16.0&--&--&--\\
91987.088&            CH$_3$CN~                                 J$_K$=5$_0$-4$_0$&10.00&  8.8&--&--&--\\
60058.127&         H$_2$CCO(o)~                                 3$_{13}$-2$_{12}$& 8.00& 15.9&-1.62( 0.99)&21.20( 2.30)&  132.(   18.)\\
61190.279&         H$_2$CCO(o)~                                 3$_{12}$-2$_{11}$& 8.00& 16.0&--&--&  130.(   18.)\\
80076.652&         H$_2$CCO(o)~                                 4$_{14}$-3$_{13}$&11.25& 18.8&--&--&   76.(   19.)\\
81586.230&         H$_2$CCO(o)~                                 4$_{13}$-3$_{12}$&11.25& 18.9&--&--&   90.(   20.)\\
65944.301&                HNCO~                                 3$_{03}$-2$_{02}$& 3.00&  3.2&-5.12( 0.88)&22.00( 2.10)&   72.(   17.)\\
87925.237&                HNCO~                                 4$_{04}$-3$_{03}$& 4.00&  6.3&--&--&  247.(   21.)\\
86846.960&                 SiO~                                             J=1-0& 2.00&  2.1&-2.29( 0.18)&20.60( 0.40)& 1262.(   22.)\\
85759.199&          $^{29}$SiO~                                             J=1-0& 2.00&  2.1&--&--&  112.(   28.)\\
57722.670&           CH$_3$CHO~                               3$_{03}$-2$_{02}$ A& 3.25&  2.8&-0.80( 0.73)&19.30( 1.70)&   99.(   25.)\\
59285.370&           CH$_3$CHO~                               3$_{12}$-2$_{11}$ E& 2.85&  5.2&--&--&   90.(   16.)\\
59379.499&           CH$_3$CHO~                               3$_{12}$-2$_{11}$ A& 2.88&  5.1&--&--&   81.(   16.)\\
74891.677&           CH$_3$CHO~                               4$_{14}$-3$_{13}$ A& 4.00&  7.7&--&--&   94.(   17.)\\
74924.134&           CH$_3$CHO~                               4$_{14}$-3$_{13}$ E& 4.00&  7.7&--&--&  122.(   17.)\\
76866.436&           CH$_3$CHO~                               4$_{04}$-3$_{03}$ E& 4.32&  5.6&--&--&   87.(   21.)\\
76878.953&           CH$_3$CHO~                               4$_{04}$-3$_{03}$ A& 4.32&  5.5&--&--&   85.(   20.)\\
79099.313&           CH$_3$CHO~                               4$_{13}$-3$_{12}$ E& 4.00&  8.1&--&--&   74.(   20.)\\
79150.166&           CH$_3$CHO~                               4$_{13}$-3$_{12}$ A& 4.00&  8.0&--&--&   89.(   20.)\\
69002.890&                  NS~                            J=3/2-1/2 F=5/2-3/2$e$& 2.00&  1.1&-2.61( 0.78)&20.90( 1.80)&  129.(   12.)\\
69017.991&                  NS~                            J=3/2-1/2 F=3/2-1/2$e$& 0.74&  1.1&--&--&   48.(    0.)\\
69037.497&                  NS~                            J=3/2-1/2 F=3/2-3/2$e$& 0.59&  1.1&--&--&   38.(    0.)\\
69040.324&                  NS~                            J=3/2-1/2 F=1/2-1/2$e$& 0.59&  1.1&--&--&   38.(    0.)\\
69059.844&                  NS~                            J=3/2-1/2 F=1/2-3/2$e$& 0.07&  1.1&--&--&    5.(    0.)\\
69283.195&                  NS~                            J=3/2-1/2 F=1/2-3/2$f$& 0.07&  1.1&--&--&    5.(    0.)\\
69330.592&                  NS~                            J=3/2-1/2 F=3/2-3/2$f$& 0.59&  1.1&--&--&   49.(    0.)\\
69411.943&                  NS~                            J=3/2-1/2 F=5/2-3/2$f$& 2.00&  1.1&--&--&  135.(   14.)\\
69437.850&                  NS~                            J=3/2-1/2 F=1/2-1/2$f$& 0.59&  1.1&--&--&   40.(    0.)\\
69485.223&                  NS~                            J=3/2-1/2 F=3/2-1/2$f$& 0.74&  1.1&--&--&   50.(    0.)\\
69746.747&          H$_2$CS(o)~                                 2$_{11}$-1$_{10}$& 4.50& 14.8&-1.60( 0.79)&14.10( 1.90)&   90.(   13.)\\
68699.385&          H$_2$CS(p)~                                 2$_{02}$-1$_{01}$& 2.00&  1.6&--&--&   67.(   11.)\\
62931.800&                  SO~                                 J$_K$=2$_1$-1$_0$& 1.92&  1.4&-1.01( 0.30)&19.70( 0.70)&  644.(   20.)\\
86093.950&                  SO~                                 J$_K$=2$_2$-1$_1$& 1.50& 15.2&--&--&   53.(   18.)\\
69408.371&              SO$^+$~                                     J=3/2-1/2 $e$& 1.31&  0.0& 1.54( 2.00)&20.00( 0.00)&   63.(   10.)\\
69783.846&              SO$^+$~                                     J=3/2-1/2 $f$& 1.31&  0.0&--&--&   63.(    0.)\\
57083.814&              C$_4$H~                                 N=6-5 J=13/2-11/2&13.00&  6.8&-1.05( 0.59)&18.50( 1.40)&  165.(   28.)\\
57122.461&              C$_4$H~                                  N=6-5 J=11/2-9/2&11.00&  6.9&--&--&   79.(   26.)\\
66600.700&              C$_4$H~                                 N=7-6 J=15/2-13/2&15.00&  9.6&--&--&  114.(   16.)\\
66639.313&              C$_4$H~                                 N=7-6 J=13/2-11/2&13.00&  9.6&--&--&  143.(   16.)\\
76117.438&              C$_4$H~                                 N=8-7 J=17/2-15/2&17.00& 12.8&--&--&  177.(   20.)\\
76156.031&              C$_4$H~                                 N=8-7 J=15/2-13/2&15.00& 12.8&--&--&  150.(   20.)\\
85634.012&              C$_4$H~                                 N=9-8 J=19/2-17/2&19.00& 16.4&--&--&   89.(   27.)\\
85672.578&              C$_4$H~                                 N=9-8 J=17/2-15/2&17.00& 16.5&--&--&   66.(   26.)\\
63686.052&             HC$_3$N~                                             J=7-6& 7.00&  9.2&-3.90( 0.46)&21.40( 0.90)&  396.(   20.)\\
72783.822&             HC$_3$N~                                             J=8-7& 8.00& 12.2&--&--&  300.(   17.)\\
81881.468&             HC$_3$N~                                             J=9-8& 9.00& 15.7&--&--&  236.(   20.)\\
90979.023&             HC$_3$N~                                            J=10-9&10.00& 19.6&--&--&  151.(   28.)\\
57229.067&              C$_2$S~                                 J$_N$=5$_4$-4$_3$& 5.00&  5.4&-1.35( 0.83)&18.20( 2.00)&   84.(   27.)\\
69281.115&              C$_2$S~                                 J$_N$=6$_5$-5$_4$& 6.00&  8.2&--&--&  142.(   14.)\\
81505.170&              C$_2$S~                                 J$_N$=7$_6$-6$_5$& 7.00& 11.5&--&--&   72.(   18.)\\
\hline \end{longtable} \tablefoot{ 
For species with hyperfine structure, the relative intensities were fixed, 
and the given uncertainty corresponds to that in the total integrated intensity. 
} \twocolumn 

\section{Notes on individual molecules} \label{Notes}

In this section, we give short reviews of the different species
and discuss constraints or new results obtained from this survey.

\subsection{C$_2$H} 

The highly reactive ethynyl radical was first discovered and identified toward the Orion
nebula by \cite{tuc74} based on its hyperfine structure, while it had still not been
observed in the laboratory. C$_2$H has turned out to be both an ubiquitous and abundant
species in the interstellar medium, including in diffuse clouds (\citealt{luc00b}) where its
abundance is of the order of a few $10^{-8}$ with respect to H$_2$.

\begin{figure}[h] \begin{center}
\includegraphics[width=8cm]{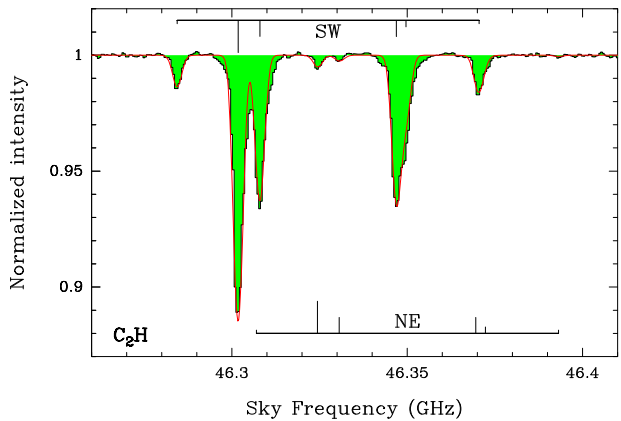}
\caption{Spectrum of the C$_2$H $N$=1-0 line. The different hyperfine components are indicated with
their relative intensities for the SW ({\em top}) and NE ({\em bottom}) absorptions. Note the clear
detection of the two strongest hyperfine components arising from FG0.89NE. The red curve shows the
LTE synthetic spectrum, taking column densities of 1.3$\times$$10^{15}$ and 4$\times$$10^{13}$~cm$^{-2}$
toward the SW and NE images respectively, and a rotation temperature of 5.14~K.}
\label{fig-c2h}
\end{center} \end{figure}

The ground state transition $N$=1-0 at $\nu_0$=87~GHz is decomposed into six components by hyperfine splitting.
Toward PKS~1830$-$211, the strongest hyperfine components were previously observed by
\cite{men99} with the VLA, although at low spectral resolution.
We detect here all six $N$=1-0 hyperfine components arising from the SW absorption, as well as
the two strongest hyperfine components (the $J$=3/2-1/2, $F$=2-1 and $F$=1-0) related to the
NE absorption at $-147$ km\,s$^{-1}$ (see Fig.\ref{fig-c2h}). The absorption depths reflect the
LTE relative intensities of the hyperfine components, indicating that the lines are optically thin.


\subsection{HCN, HNC and HCO$^+$}

\begin{figure}[h] \begin{center}
\includegraphics[width=8cm]{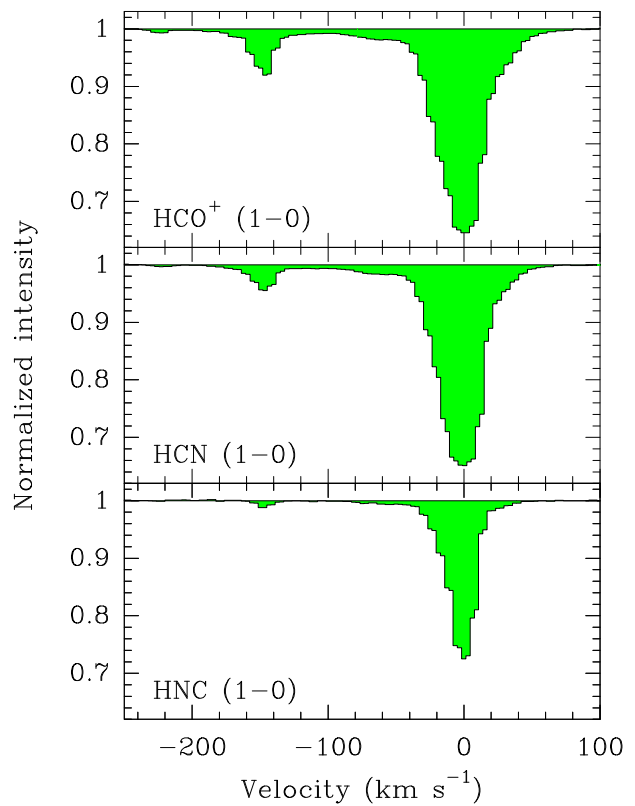}
\caption{Full spectra of the HCO$^+$, HCN, and HNC 1-0 transitions, showing the absorption toward the
SW (0~km\,s$^{-1}$) and NE ($-$147~km\,s$^{-1}$) images of the quasar. A zoom in with a narrower intensity
scale, shown in Fig.\ref{fig-newvelocomp}, reveals additional velocity components.}
\label{fig-strongabs}
\end{center} \end{figure}

Molecular absorption was first reported toward PKS~1830$-$211 by \cite{wik96b} after the detection of
the HCN, HNC, and HCO$^+$ 2-1 transitions, allowing the redshift of the intervening galaxy to be derived.
Fig.\ref{fig-strongabs} shows the full spectra of the HCO$^+$, HCN, and HNC 1-0 transitions observed in
our survey. The first two lines are likely saturated and their opacities could be underestimated.
In particular two findings seem to indicate that the HCO$^+$ column density might be underestimated by a factor of 2-3.
First, the HCO$^+$/H$^{13}$CO$^+$ ratio toward FG0.89SW is a factor of $\sim$2 lower than 
the corresponding HCN/H$^{13}$CN and HNC/HN$^{13}$C ratios. Second, the ratio of
the column densities along the two lines of sight, 
SW/NE is also a factor 2-3 lower for HCO$^+$ than for other species (see Table~\ref{tab-NE}).
In any case, the uncertainty in the opacity introduced by the uncertainty in the continuum background
illumination is nevertheless large for both HCO$^+$ and HCN (see \S.\ref{back-illumin})

Time variations and the discovery of additional velocity components are discussed in
\S.\ref{section-timvar} and \S.\ref{section-newvelocomp}, respectively.

\subsection{N$_2$H$^+$} 

N$_2$H$^+$ is thought to be one of the most abundant interstellar ions, and plays a key role in nitrogen
chemistry. It is detected in a large number of Galactic clouds. However, it remains elusive in diffuse clouds
(\citealt{lis01}), with an abundance upper limit more than two orders of magnitude lower than its abundance
in FG0.89SW. We note that N$_2$H$^+$ is also rare in translucent clouds, e.g., \cite{tur95a} having detected
it in only 2 out of 16 such clouds.


\subsection{HOC$^+$}

The hydroxymethylidynium ion, HOC$^+$, isomer of the formyl ion, was first tentatively identified in Sgr~B2 by
\cite{woo83}. Its presence in several dense molecular clouds was then confirmed by \cite{ziu95} and \cite{app97},
with a relative abundance HCO$^+$/HOC$^+$$\sim$360--6000. This abundance ratio is however about one order of
magnitude lower ($\sim$70--120) in diffuse clouds (\citealt{lis04}), and seemingly has an in-between value in PDR:
\cite{app99} derived HCO$^+$/HOC$^+$$\sim$270 toward the Orion bar. \cite{mar09} find a ratio $\sim$70 toward NGC253,
similar to that in diffuse clouds.

In FG0.89SW, we derive a ratio HCO$^+$/HOC$^+$ of $\sim$60, although this value might be slightly underestimated
because of the large opacity of HCO$^+$. The H$^{13}$CO$^+$/HOC$^+$ ratio is indeed $2.6 \pm 0.1$. Assuming a
$^{12}$C/$^{13}$C ratio of 35, the HCO$^+$/HOC$^+$ ratio in FG0.89SW is comparable to that in diffuse clouds.



\subsection{HCO}

The formyl radical, HCO, was discovered in the interstellar medium by \cite{sny76} and has been identified
as a tracer of the cloud-illuminated interface in PDRs (\citealt{dejon80, sch01}).
\cite{gar02} revealed widespread HCO emission in the nuclear starburst of M82, with HCO, CO, and HII emission
in nested rings, HCO extending farther out in the disk. \cite{ger09} demonstrated that the
HCO emission in the Horsehead nebula traces the PDR front, coincident with PAH and hydrocarbon emission.
The abundance of HCO in the PDR is comparable to that of HCO$^+$, whereas it becomes about one order of
magnitude less in the dense regions shielded from UV radiation. To the best of our knowledge, HCO has not
been detected in diffuse or translucent clouds to date, although it is located close in frequency to the
H$^{13}$CO$^+$ (1-0) and SiO (2-1) lines.

We clearly detect three (out of the four) hyperfine components of the ground state transition $1_{01}-0_{00}$,
the remaining low intensity component being buried in the noise level. The relative optical depths are
consistent with LTE intensities, suggesting that the lines are optically thin. We derive an abundance 
ratio HCO$^+$/HCO of $\sim$13, similar to the values observed in dense shielded regions.
This value is even larger if we consider that the HCO$^+$ opacity might be underestimated by a factor 2--3.


\subsection{CH$_2$NH} 

We identify the absorption feature near 33.93 GHz with the $1_{01}-0_{00}$ ground state transition
of methanimine, as it is the simplest interstellar molecule with lowest energy transition close to
that frequency. The hyperfine structure is not resolved. Only one line of methanimine is located
within the limits of our frequency coverage.
The strongest transitions, with predicted $\tau > 0.1$ based on the
$1_{01}-0_{00}$ absorption depth, have rest frequencies of 127.9, 166.9, 225.6 and 284.3 GHz.
Of these, only the second and last fall in an ALMA band for a redshift of $z$=0.89.

Methanimine was first detected in Sgr~B2 by \cite{god73} and has been further observed in
several star-forming regions and translucent clouds (\citealt{tur99}),
but is not detected in cold dark clouds (\citealt{tur91,dic97}).
It has been tentatively detected in NGC253 (\citealt{mar06}) and has been observed
in Arp~220 (\citealt{sal08}). Methanimine is a prebiotic molecule, precursor to more complex
molecules such as glycine.

\subsection{CH$_3$NH$_2$} 

Methylamine is the saturated and terminal product of the hydrogenation series based on the
cyanide radical: CN-HCN-CH$_2$NH-CH$_3$NH$_2$.

Methylamine has a very complex spectrum resulting from the combination of internal rotation
of the CH$_3$ group, the asymmetry of the molecule and the inversion of the NH$_2$ group
(\citealt{tak73,ily07}). It was first detected in Sgr~B2 by \cite{kai74} (see also \citealt{tur91}).
No extragalactic detection has been reported to date.


\subsection{CH$_3$OH} 

Interstellar methanol was first detected toward Sgr~B2 and Sgr~A by \cite{bal70} and is widely
observed toward star forming regions, dark clouds, and also translucent clouds (\citealt{tur98}).
It has been detected in various extragalactic sources (e.g. \citealt{hen87}).
Methanol is about one order of magnitude less abundant in cold/translucent clouds than in
star forming regions, where its abundance can reach a few 10$^{-8}$ relative to H$_2$.
It is not detected in diffuse clouds, down to upper limits of a few lower than its abundance
in cold clouds (\citealt{lis08}). Methanol is the terminal product of hydrogenation of CO.

The methanol spectrum has a multitude of transitions, of which only one is strong enough to be
detected within our frequency coverage. Therefore, no excitation analysis can be
deduced. Assuming LTE equilibrium with the CMB radiation field, we derive a relatively high methanol
abundance of a few 10$^{-8}$ relative to H$_2$ in FG0.89SW. Several strong lines (with predicted
$\tau$$\sim$0.02--0.4) near rest frequency of 300 GHz and redshifted in the 2~mm band are
(simultaneously) observable (e.g. with ALMA or PdBI) and could provide a stringent measurement
of the methanol rotation temperature.

The calculations of \cite{jan11} have shown that methanol could be a sensitive probe of spatial
and temporal variations in the proton-to-electron mass ratio.

\subsection{l-C$_3$H} 

\begin{figure}[h] \begin{center}
\includegraphics[width=8cm]{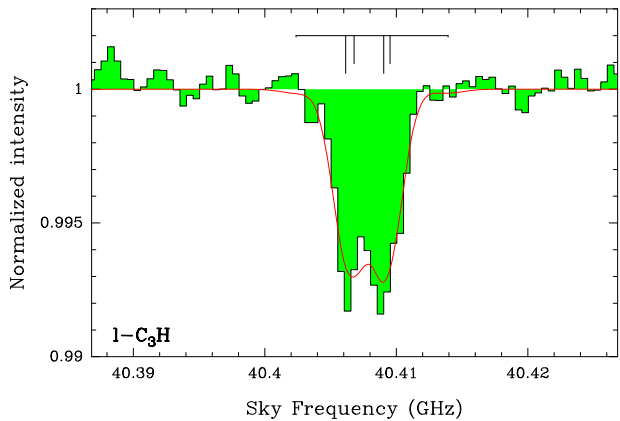}
\caption{Spectrum of the l-C$_3$H $J$=7/2-5/2 transition. The locations of the two $\Lambda$-doublets with
hyperfine structure are indicated.}
\label{fig-l-c3h}
\end{center} \end{figure}

We identify the feature around 40.4 GHz with the $J$=7/2-5/2 transition of the $^2\Pi_{1/2}$ rotational
ladder of the propynylidyne radical ($\nu_0$=76.2~GHz). The transition is split into two $\Lambda$-doublets
(see Fig.\ref{fig-l-c3h}), approximately equal in strength and spaced by about 5 MHz in the rest frame,
which secures our identification. The hyperfine structure is unresolved. No other transitions of l-C$_3$H
are detected in our frequency coverage.

Linear propynylidyne was first identified by \cite{tha85a} toward IRC+10216 and TMC~1. It is observed
in translucent clouds (\citealt{tur00b}) but remains undetected in diffuse clouds (\citealt{luc00b}, although
this might just be a sensitivity issue, as their upper limits of l-C$_3$H/HCO$^+$ is still higher
than the value we derive). No extragalactic detections have been reported to date that we are aware of.

The cyclic isomer form c-C$_3$H is detected in dense (\citealt{yam87,fos01}) and translucent (\citealt{tur00b})
clouds, with c-C$_3$H/l-C$_3$H ratio $\sim$5. No transitions of c-C$_3$H are detected in our survey.
In particular, the non-detection of the $N_{KaKc}$=2$_{12}$-1$_{11}$ transition at a rest-frame frequency of 91.5 GHz,
which is connected to the ground state, yields an upper limit of $\sim$3$\times$10$^{12}$~cm$^{-2}$, and
a c-C$_3$H/l-C$_3$H ratio $<$0.4.
We note that a low ratio c-C$_3$H/l-C$_3$H$<1$ was found by \cite{tur00b} toward one translucent cloud
(CB228).


\subsection{c-C$_3$H$_2$} 

The ring molecule cyclo-propenylidene c-C$_3$H$_2$ is a fairly ubiquituous interstellar species
(e.g., \citealt{tha85b,mat85b,cox88,mad89}).

We detect 2 ortho and 2 para lines, yielding an ortho/para ratio of $2.8 \pm 0.1$, comparable to the
statistical weight 3:1.

A tight correlation exists between the abundances of C$_2$H and c-C$_3$H$_2$ in diffuse clouds
(\citealt{luc00b,ger11}), with $N$(C$_2$H)=($28\pm 1.4$)$N$(c-C$_3$H$_2$).
We derive a comparable abundance ratio C$_2$H/c-C$_3$H$_2$=$23.6 \pm 0.2$.

\begin{figure}[h] \begin{center}
\includegraphics[width=8cm]{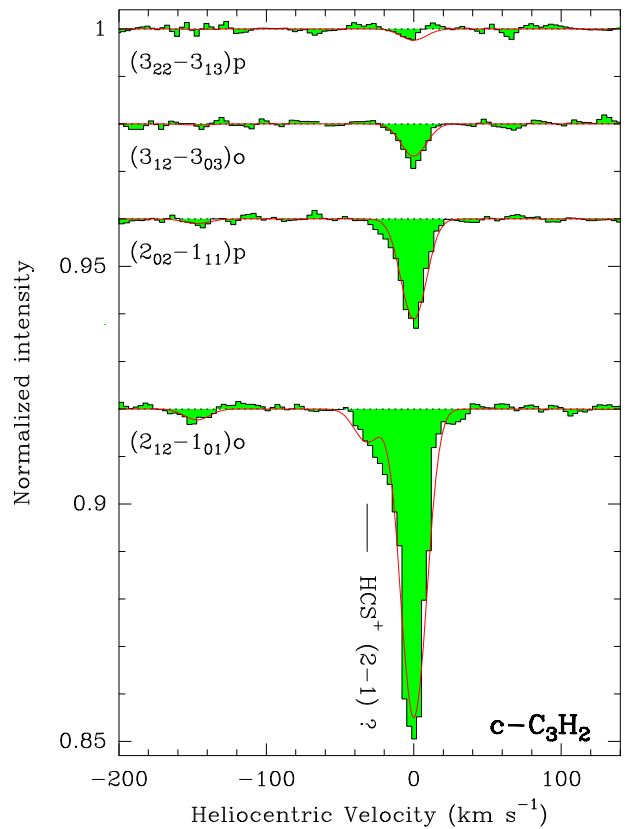}
\caption{c-C$_3$H$_2$ transitions. The dashed curve on the wing of the $2_{12}-1_{01}$ line
corresponds to the LTE prediction for the HCS$^+$ $J$=2-1 line with a HCS$^+$ column
density of $5 \times 10^{12}$ cm$^{-2}$.}
\label{fig-c-c3h2}
\end{center} \end{figure}

The excess absorption in the blue wing of the c-C$_3$H$_2$ $2_{12}$-1$_{01}$ line (see
Fig.\ref{fig-c-c3h2}) could be due to the known velocity component at $\sim$$-20$~
km\,s$^{-1}$ toward FG0.89SW (see e.g. Fig.5 from \citealt{mul06}). Alternatively, we
note that this feature could be due the HCS$^+$ 2-1 line, which would fall at
$\sim$$-32$~km\,s$^{-1}$ with respect to the center of the c-C$_3$H$_2$ $2_{12}$-1$_{01}$
line. HCS$^+$ is indeed detected in diffuse clouds with
an abundance of $\sim$3$\times$10$^{-10}$ relative to H$_2$ (\citealt{luc02}). If its
abundance were similar in FG0.89SW, then the LTE predicted absorption depth would be of the
order of $\tau$=0.015, i.e. compatible with the observations. There are no other lines
of HCS$^+$ within our frequency coverage. Observations of the $J$=1-0, 3-2, or 4-3
transitions (redshifted to 22.6, 67.9, and 90.5 GHz, and with predicted opacity of
$\tau$=0.007, 0.013, and 0.006, respectively) could confirm the presence of HCS$^+$.

\subsection{l-C$_3$H$_2$} 

Propadienylidene, l-C$_3$H$_2$, is the linear isomer of c-C$_3$H$_2$ that was first discovered by
\cite{cer91} in the carbon-chain rich dark cloud TMC~1. Noticeably, \cite{mai11} identified
l-C$_3$H$_2$ as the probable carrier of two conspicuous diffuse interstellar bands
(at 4881 and 5450 \AA).

The ortho/para ratio is $2.4_{-0.5}^{+1.0}$, in agreement with the statistical weight of 3:1.

The linear form is usually less abundant than the cyclic one, with a cyclic/linear ratio of 3--5
in diffuse clouds, 70 in TMC~1, and $>$150 in Sgr~B2 (\citealt{cer91,cer99}, see also \citealt{tur00b}).
We derive a cyclic/linear ratio of $25 \pm 3$, which is in-between the ratio observed toward diffuse
clouds and TMC~1.








\subsection{H$_2$CCN} 

The non-linear cyanomethyl radical has a very complicated spectrum owing to the unpaired electron and 
hyperfine splitting induced by the H and N atoms.
We detect two absorption features, corresponding to $N_{K_a,K_b}$=$3_{03}$-$2_{02}$ and $4_{04}$-$3_{03}$.
Both are from the ortho species (ortho states are characterized by $K_a$ even and para states by $K_a$ odd).


\cite{irv88} first identified H$_2$CCN toward Sgr~B2 and TMC~1.
It was also detected in the circumstellar envelope of IRC+10216 (\citealt{agu08}),
but, to the best of our knowledge, it has not been observed toward other sources.
In particular, this is the first extragalactic detection of this molecule.


\subsection{CH$_3$CCH} 

Methyl acetylene was first detected in Sgr~B2 and Orion A by \cite{sny73}.
It is observed in dark clouds (e.g., \citealt{irv81}) and has been detected in the nearby starburst galaxies
NGC253, NGC4945 and M82 (e.g. \citealt{mau91,wan04}).

We detect the $J_K$=5$_0$-4$_0$ transition, but not the $J_K$=4$_0$-3$_0$, suggesting that the
rotation temperature is higher than 5.14~K. The kinetic temperature was fixed to 50~K (see \ref{ch3cn}).

We note that CH$_3$CCH is the molecule in our survey which shows the miminum scatter in relative abundances
between the different sources.


\subsection{CH$_3$CN} \label{ch3cn} 


Methyl cyanide was first detected in space toward the Galactic center by \cite{sol71}. It is commonly detected
in hot cores (e.g. \citealt{olm93}), where the heating from massive stars is thought to evaporate the products
of grain surface chemistry. The abundance of CH$_3$CN in these regions of active star formation exceeds $10^{-9}$
relative to H$_2$ (e.g. \citealt{kal00}). We note that CH$_3$CN is also observed in dark clouds (\citealt{mat83}) and has been
detected in NGC253 (\citealt{mau91}). On the other hand, CH$_3$CN is detected in neither diffuse (\citealt{lis01})
nor translucent clouds (\citealt{tur00a}), with upper limits clearly lower than abundances in dark clouds.

Symmetric-top molecule are excellent temperature probes, because radiative transitions between
the different $K$-ladders are forbidden, so that only collisional excitation is effective.
Consequently, the rotation temperature of the molecule can be derived within the same $K$-ladder
on the one hand, and the kinetic temperature is reflected in the population
of the $K$ levels on the other hand (e.g. \citealt{chu83}).

In our LTE synthetic spectrum approach, the term $\exp{(\frac{E_l}{k_BT_{rot}})}$ in Eq.\ref{eq-ncol} was changed to
$\exp{\left[ \frac{(E_l-E_{KK})}{k_BT_{rot}} \right ]}\exp{\left [ \frac{E_{KK}}{k_BT_{kin}} \right ]}$,
where $E_{KK}$ is the lowest energy level of each $K$-ladder,
to evaluate the opacity of each transition (see e.g., \citealt{olm93}).
The partition function was calculated accordingly by taking into account both the rotation and kinetic temperatures
\begin{equation}
Q = \sum_{JK} g_Ig_Kg_J
\exp{\left ( \frac{-(E_l-E_{KK})}{k_BT_{rot}} \right )} \exp{\left ( \frac{-E_{KK}}{k_BT_{kin}} \right )},
\end{equation}
\noindent where $g_J$=(2$J$+1) and the product $g_Ig_K$=2 for $K$=0 and $K$$\ne$3n (n$>$0), and $g_Ig_K$=4 for $K$=3n.

\begin{figure}[h] \begin{center}
\includegraphics[width=8cm]{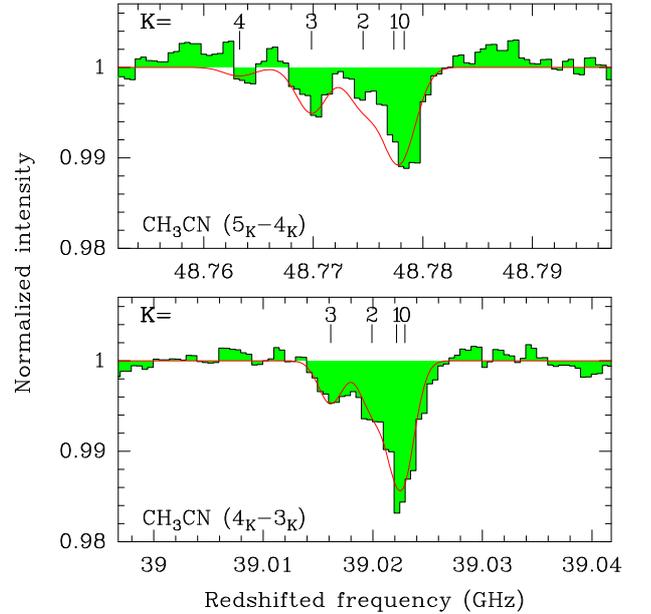}
\caption{Spectra of the $4_K-3_K$ and $5_K-4_K$ transitions of CH$_3$CN. Our LTE prediction
assuming a kinetic temperature of 50~K is overlaid.}
\label{fig-ch3cn}
\end{center} \end{figure}

Two groups of transitions of methyl cyanide fall in the frequency coverage of our survey, the
$J_K$=$4_K$-$3_K$ redshifted near 39~GHz, and the $J_K$=$5_K$-$4_K$, redshifted to $\sim$48.8~GHz. Both
are clearly detected. Remarkably, the $K$-ladder transitions are detected up to $K$=3 (see Fig.\ref{fig-ch3cn}),
with lower energy state up to 70~K above the ground level. We note that CH$_3$CN is the only species in our survey
showing absorption from such high energy levels. This suggests that the kinetic temperature is high,
in agreement with the observations of inversion lines of ammonia up to $(J,K)$=10,10
(which lies at about 1000~K above the ground state) by \cite{hen08}. On the basis of their estimates, the kinetic
temperature $T_{kin}$ is about 80~K for 80-90\% of the ammonia gas, while the remaining fraction could even
reach $T_{kin}$$>$600~K.

A robust fitting of the CH$_3$CN spectra is difficult as the different $K$-ladder transitions are close
in frequency relative to the width of the lines.
Nevertheless, a good match of the observations could be found using our LTE model by assuming a CH$_3$CN column
density of 1.3$\times$$10^{13}$ cm$^{-2}$ and a kinetic temperature of 50~K (see Fig.\ref{fig-ch3cn}).
This latter parameter value is comparable to, although slightly lower than that obtained by \cite{hen08}.
We neglect the broadening of the line due to hyperfine splitting, as it is at most of 1.5~MHz for the principal
hyperfine components of the $J_K$=$5_4$-$4_4$ transition.



\subsection{H$_2$CCO} 

Ketene was first detected in Sgr~B2 by \cite{tur77}. It was detected in line surveys of
regions of low-mass and massive star formation (e.g. \citealt{tur91,vandis95,rui07}), as well
as in both translucent and dark clouds (\citealt{tur99,kai04}), suggesting that
it is a common interstellar species. 
However, only few specific studies of the molecule have
been conducted, and its formation mechanism(s) is subject to controversy: H$_2$CCO could be
either produced by gas-phase reactions (\citealt{tur99}) or formed by means of surface chemistry,
followed by grain desorption (\citealt{rui07}). Its typical abundance is of the order of
$10^{-10}$--$10^{-9}$ relative to H$_2$.


We detect four different (ortho) transitions of H$_2$CCO in our survey, yielding
a rotation temperature of $2.6\ _{-0.8}^{+1.1}$~K, significantly lower than $T_{CMB}$=5.14~K at $z$=0.89.


\subsection{HNCO} 

Isocyanic acid was first discovered toward Sgr~B2 (\citealt{sny72}).
We detect two lines of HNCO, the 3$_{03}$-2$_{02}$ and the 4$_{04}$-3$_{03}$, the latter in a low-noise
region of the survey.
They do not appear to be at LTE equilibrium, and we could not derive a positive rotation temperature.
The centroid velocity of HNCO is also significantly offset with respect to the average velocity of all molecules.
We could not identify lines from other molecules close in frequency, and therefore discard the possibility of
a false identification.






\subsection{SiO} 

Silicon monoxide is another commonly observed interstellar molecule, considered as a good shock tracer.
Its abundance is low in cold dark clouds (of the order of or less than $10^{-12}$ relative to H$_2$),
where it is presumably heavily depleted onto dust grains, but can reach values several orders of magnitude
higher in outflows and shocked gas. It is found to have an intermediate abundance (SiO/H$_2$$\sim$$10^{-10}$)
in diffuse and translucent clouds (\citealt{luc00a,tur00a}). Toward FG0.89SW, we find a SiO abundance of
nearly 4$\times$10$^{-10}$ that is comparable to, though slightly higher than, that in diffuse/translucent clouds.

Interestingly, \cite{ziu89} found a correlation between ln[SiO/HCN] and the inverse of the kinetic temperature in
Galactic molecular clouds: were this correlation to hold for the gas in FG0.89SW, the SiO/HCN ratio would then point
to a kinetic temperature of $\sim$30~K, a value not so far from the kinetic temperature estimated from CH$_3$CN.

The $^{29}$SiO isotopologue is also detected and we derive a $^{28}$Si/$^{29}$Si ratio of $11_{-2}^{+4}$, which is
roughly half the Solar System value.


\subsection{CH$_3$CHO} 

Acetaldehyde is an asymmetric rotor, with two symmetry states (A and E) resulting from the
rotation of the methyl group. Electric dipole transitions are not allowed between these
states, which have the same weight statistic. Each transition is dedoubled by the symmetry, and the
resulting spectrum consists of a series of closeby A/E doublets.

The detection of acetaldehyde was first reported toward Sgr~B2 (\citealt{bal71}) and the molecule 
has been observed in cold dark clouds (\citealt{mat85a}) and translucent clouds (\citealt{tur99}).
Interestingly, \cite{che03} observed that CH$_3$CHO is more widespread toward the Galactic center
than most other complex organic molecules. They argue that the molecule could be produced by
shock-induced disruption on dust mantles.

No extragalactic detection has yet been reported for this species.


\subsection{NS} 

Nitrogen sulfide was first detected in Sgr~B2 by \cite{got75}. It is detected in quiescent clouds
(\citealt{mcg94}) and in massive-star forming regions (\citealt{mcg97}). There are no reports of
NS detection in diffuse or translucent clouds to date.

\begin{figure}[h] \begin{center}
\includegraphics[width=8cm]{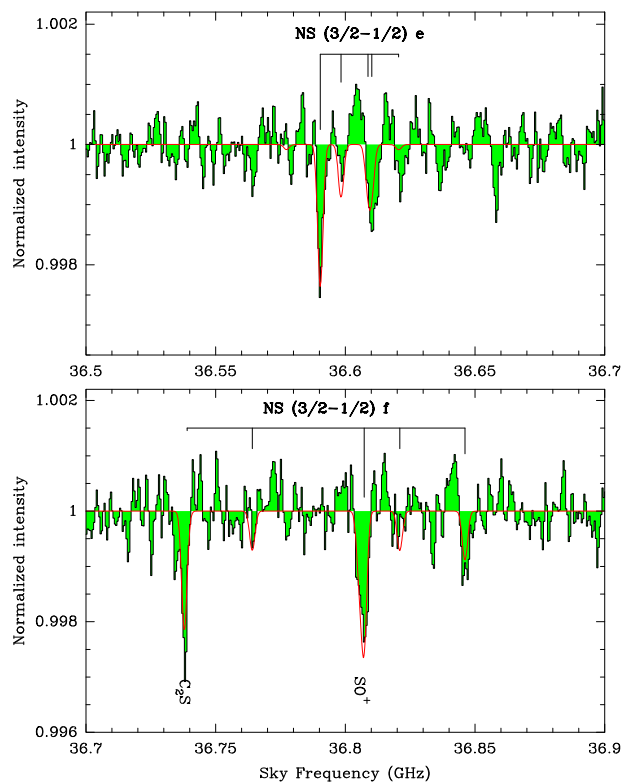}
\caption{Spectrum of NS (2-1). The hyperfine structure with relative intensities is indicated.}
\label{fig-ns}
\end{center} \end{figure}

The rotational levels of the molecule are divided into two levels ($e$/$f$) with opposite parity by
$\Lambda$-doubling, each of them subdivided into hyperfine structure corresponding to the nitrogen nucleus.
We clearly detect the hyperfine components with highest relative intensities (see Fig.\ref{fig-ns}).
We note that the frequency interval between $\sim$36 and 40~GHz is the region observed with the lowest
noise level in our survey, allowing these weak lines of NS, H$_2$CS, C$_2$S, and SO$^+$ to be detected.

Taking the CS column density estimated by \cite{mul06}, we find a ratio NS/CS$\sim$0.02, comparable
to the values 0.02--0.05 measured by \cite{hat02} in a sample of hot cores. \cite{mar03} estimated
a ratio NS/CS=0.4 in NGC253, which they interpreted as the signature of NS chemical enhancement by
shocks (see also \citealt{vit01}).

\subsection{H$_2$CS} 

Thioformaldehyde was first detected in absorption toward Sgr~B2 by \cite{sin73}.
It is observed in star forming regions and cold dark clouds (e.g. \citealt{irv89,min91})
as well as in translucent clouds (\citealt{tur96}).
Tentative extragalactic detections are reported by \cite{hei99} in the LMC and \cite{mar06} in NGC253.

Two lines of H$_2$CS are detected, one of each ortho and para form, from which we derive an ortho/para
ratio of $3.0_{-0.5}^{+0.8}$.
However, we note that another ortho line, the 2$_{12}$-1$_{11}$ at a rest frequency of 67.653820 GHz, 
should have been detected at the same intensity as the 2$_{11}$-1$_{10}$ line, according to the LTE prediction.
It is fitted with an opacity twice lower, although with a significance of just 3~$\sigma$ on its integrated
opacity, which is the reason we discarded it from our estimate of the ortho-H$_2$CS column density.

The H$_2$CO/H$_2$CS ratio, derived from their para forms, is $32 \pm 6$, fully consistent with the
present oxygen-to-sulphur cosmic abundance ratio (33:1, \citealt{lod03}).

\subsection{SO} 

Sulfur monoxide was first detected by \cite{got73} toward several star forming regions
and later observed in dark clouds by \cite{ryd80}.

From the $J_K$=$2_1$-$1_0$ line detected in our survey, we derive an abundance of $\sim$$10^{-9}$ relative to H$_2$,
similar to that observed in Galactic diffuse clouds (\citealt{luc02}), and $\sim$25 lower than that in translucent
clouds (\citealt{tur95b}).

The non-detection of the SO$_2$ 1$_{11}$-0$_{00}$ ground state transition yields a ratio SO/SO$_2$$>$8.
Both species are observed in quiescent translucent clouds with an abundance ratio SO/SO$_2$$\sim$1--15 (\citealt{tur95b}).
It has been found that SO$_2$ is not detected in diffuse clouds, down to a fractional abundance $<$10$^{-9}$ (\citealt{luc02}).

The non-detection of $^{34}$SO points to an upper limit of SO/$^{34}$SO$>$13.7 (3$\sigma$). \cite{mul06} derived
a ratio $^{32}$S/$^{34}$S of $10.5 \pm 0.6$ from measurements of CS and H$_2$S, and their isotopologues.





\subsection{SO$^+$} 

The reactive radical SO$^+$ is characterized by a $^2$$\Pi$$_{1/2}$ ground state. Its spin-doubled rotational
transitions are split by $\Lambda$-type doubling.

The SO$^+$ $J$=3/2-1/2~$e$ line is partly blended with the $F$=5/2-3/2~$f$ hyperfine transition of NS.
The width of the spectral feature at $\nu_{sky}$$\sim$36.807~GHz, as obtained by the fit
of a single Gaussian component, is $37 \pm 5$~km\,s$^{-1}$, i.e. much larger than the width of other lines.
Nevertheless, after taking into account the SO$+$ $e$/$f$ doublet together with the NS line, the LTE synthetic
spectrum reproduces fairly well the observations.

It has been found that SO$^+$ is a rather widespread and quite abundant interstellar species (\citealt{tur94}), 
and is seen, although weakly, in translucent clouds (\citealt{tur96}). There, the 
abundance ratio SO/SO$^+$ is of the order of 30.
We estimate a ratio SO/SO$^+$=$15 \pm 3$ of comparable order of magnitude toward FG0.89SW.

\subsection{C$_4$H} 

The butadiynyl radical was first reported by \cite{gue78} toward IRC+10216 and has been
observed toward dark clouds by \cite{irv81}. It is present in translucent clouds with a quite
high abundance of a few 10$^{-8}$ relative to H$_2$ (\citealt{tur00a}). No extragalactic detection
has been reported so far.

The interaction of the unpaired electron with the molecular rotation produces fine structure, which is
divided further into (here unresolved) hyperfine structure by the proton spin.
The resulting spectrum of C$_4$H consists of a series of closeby doublets, of which we detect four
($N$=6-5 to $N$=9-8) within the limits of our frequency coverage.



\subsection{HC$_3$N} \label{appendix-HC3N} 

The highly unsaturated cyanoacetylene molecule was first detected in Sgr~B2 (\citealt{tur71}) and
is observed in various sources, such as translucent clouds, dense dark clouds, and star forming
regions, with abundances of $\sim$10$^{-10}$--10$^{-9}$ relative to H$_2$. HC$_3$N emission has also
been detected in several extragalactic objects (see \citealt{lin11} for a recent review).

We detect all four transitions of HC$_3$N within the frequency range of our survey, from $J$=7-6
to $J$=10-9. The $J$=3-2 and $J$=5-4 lines of HC$_3$N were previously detected with the VLA by
\cite{men99}. More recently, \cite{hen09} reported the detection of all transitions from $J$=3-2
to $J$=7-6 with the Effelsberg 100~m telescope. The $J$=10-9 line was not detected. The results
of their observations are reported in the rotation diagram together with our data points
(Fig.\ref{fig-rotdiag}). They appear to be inconsistent with one another. The integrated opacities
of the $J$=7-6 line, observed in common with both works, differ by a factor of $\sim$2, as for
their fitted linewidths. Time variablity and/or errors in the spectral baseline could explain the
observed differences. We note that the higher S/N $J$=7-6 line 
appears slightly asymmetrical, with a shoulder on the blue-shifted side, reminescent of that of the
HCO$^+$ 2-1 profile observed at higher spectral resolution (\citealt{mul06,mul08}). This transition
was therefore not included in the Gaussian fit to derive the velocity and linewidth.

The points corresponding to the four lines observed by us appear remarkably well-aligned in the
rotation diagram, suggesting that the lines are at LTE. They yield a rotation temperature of
$T_{rot}$=$6.3 \pm 1.3$~K, just slightly higher than the expected CMB temperature. On the other hand,
the points corresponding to the lines observed by \cite{hen09} at lower frequencies are not aligned
in the rotation diagram. While we cannot discard possible time variations, another explanation
could be a change in the source covering factor with frequency, as discussed by \cite{hen09}.
Observations of all these transitions within a short period of time would help clarify the
interpretation of these data.

\subsection{C$_2$S} 

The thioxoethenylidene radical was first identified by \cite{sai87} in previous spectra of
Sgr~B2 and TMC~1, based on laboratory spectroscopy studies. It has been found that C$_2$S is present in
translucent clouds (\citealt{tur98+}) with a typical abundance of $\sim$10$^{-9}$ and is
detected in the 2 mm survey of NGC253 by \cite{mar06}. It is the heaviest species
detected in our survey. We measure an abundance of $\sim$2$\times$10$^{-10}$ relative to H$_2$.


Few lines ($10 \ge J_u \ge 16$) of C$_3$S fall within the frequency coverage of our survey.
Their non-detection gives a (not tightly constraining) ratio C$_2$S/C$_3$S$>$0.7.
Search for lower $J$ transitions of C$_3$S would be more sensitive.
\cite{fue90} measured a typical abundance ratio C$_2$S/C$_3$S=5 in dark clouds.


\end{appendix}

\end{document}